\documentclass[pra,aps,preprint,amsmath,amssymb,showpacs]{revtex4}
\usepackage{graphicx}
\usepackage{epsfig}
\usepackage{epstopdf}
\include{graphics}
\begin{document}

\title{Stabilizing soliton-based multichannel transmission with 
frequency dependent linear gain-loss}

\author{Debananda Chakraborty$^{1}$, Avner Peleg$^{2}$, and Quan M. Nguyen$^{3}$}

\affiliation{$^{1}$Department of Mathematics, New Jersey City University, 
Jersey City, New Jersey, 07305, USA}

\affiliation{$^{2}$Department of Basic Sciences, Afeka College of Engineering, 
Tel Aviv, 69988, Israel}

\affiliation{$^{3}$Department of Mathematics, International University, 
Vietnam National University-HCMC, Ho Chi Minh City, Vietnam}


\begin{abstract}
We report several major theoretical steps towards realizing stable long-distance 
multichannel soliton transmission in Kerr nonlinear waveguide loops. 
We find that transmission destabilization in a single waveguide 
is caused by resonant formation of radiative sidebands and 
investigate the possibility to increase transmission stability
by optimization with respect to the Kerr nonlinearity coefficient $\gamma$.  
Moreover, we develop a general method for transmission stabilization, 
based on frequency dependent linear gain-loss in 
Kerr nonlinear waveguide couplers, and implement it 
in two-channel and three-channel transmission. 
We show that the introduction of frequency 
dependent loss leads to significant enhancement of 
transmission stability even for non-optimal $\gamma$ values 
via decay of radiative sidebands, which takes place as a dynamic phase transition. 
For waveguide couplers with frequency dependent linear gain-loss, 
we observe stable oscillations of soliton amplitudes 
due to decay and regeneration of the radiative sidebands.  
\end{abstract}
\maketitle


\section{Introduction}
\label{Introduction}
The rates of transmission of information in broadband optical waveguide systems  
can be significantly increased by transmitting many pulse sequences 
through the same waveguide \cite{Agrawal2001,Mollenauer2006,Iannone98}.  
This is achieved by the wavelength-division-multiplexing (WDM) method, 
where each pulse sequence is characterized by the central frequency of its pulses, 
and is therefore called a frequency channel.    
Applications of these WDM or multichannel systems include 
fiber optics communication lines \cite{Agrawal2001,Mollenauer2006,Iannone98}, 
data transfer between computer processors through silicon waveguides \cite{Agrawal2007a,Gaeta2008}, 
and multiwavelength lasers \cite{Zhang2009,Liu2013}. 
Since pulses from different frequency channels 
propagate with different group velocities, interchannel pulse collisions are very 
frequent, and can therefore lead to severe transmission degradation \cite{Agrawal2001}. 
Soliton-based transmission is considered to be advantageous compared with other 
transmission formats, due to the stability and shape-preserving properties of the 
solitons, and as a result, has been the focus of many studies \cite{Agrawal2001,Mollenauer2006,Iannone98}. 
These studies have shown that effects of Kerr nonlinearity 
on interchannel collisions, such as cross-phase modulation and four-wave-mixing,  
are among the main impairments in soliton-based 
WDM fiber optics transmission. Furthermore, various methods for mitigation 
of Kerr-induced effects, such as filtering and dispersion-management, have 
been developed \cite{Mollenauer2006,Iannone98}. 
However, the problem of achieving stable long-distance propagation 
of optical solitons in multichannel Kerr nonlinear waveguide loops 
remains unresolved. The challenge in this case stems from two factors. 
First, any radiation emitted by the solitons stays in the waveguide loop, 
and therefore, the radiation accumulates. Second, the radiation 
emitted by solitons from a given channel at frequencies of 
the solitons in the other channels undergoes unstable growth 
and develops into radiative sidebands. Due to radiation accumulation 
and to the fact that the sidebands form at the frequencies of the propagating 
solitons it is very difficult to suppress the instability. In the current paper, 
we report several major steps towards a solution of this important problem.

In Refs. \cite{NP2010,PNC2010,PNT2015,PC2012,CPJ2013,NPT2015}, 
we studied soliton propagation in Kerr nonlinear waveguide loops 
in the presence of dissipative perturbations 
due to delayed Raman response and nonlinear gain-loss. 
We showed that transmission stabilization can be realized at short-to-intermediate distances, 
but that at large distances, the transmission becomes unstable, and the 
soliton sequences are destroyed. Additionally, in Ref. \cite{PNT2015}, 
we noted that destabilization is caused by resonant formation 
of radiative sidebands due to cross-phase modulation. 
However, the central problems of quantifying the dependence of transmission 
stability on physical parameter values and of developing general methods  
for transmission stabilization against Kerr-induced effects were not addressed. 
In the current paper we take on these problems for two-channel and 
three-channel transmission by performing extensive simulations 
with a system of coupled nonlinear Schr\"odinger (NLS) equations. 
We first study transmission in a single lossless waveguide 
and investigate the possibility to increase transmission stability 
by optimization with respect to the value of the Kerr nonlinearity coefficient. 
We then demonstrate that significant enhancement of transmission stability 
can be achieved in waveguide couplers with frequency dependent linear loss and gain 
and analyze the stabilizing mechanisms. This stabilization is realized without 
dispersion-management or filtering.

\section{The coupled-NLS propagation model}
\label{CNLS}
We consider propagation of $N$ sequences of optical pulses in an optical waveguide 
in the presence of second-order dispersion, Kerr nonlinearity, 
and frequency dependent linear gain-loss. 
We assume a WDM setup, where the pulses in each sequence 
propagate with the same group velocity and frequency, but where 
the group velocity and frequency are different for pulses from different sequences. 
The propagation is then described by the following system 
of $N$ coupled-NLS equations \cite{Agrawal2001,PNT2015}: 
\begin{eqnarray} 
i\partial_z\psi_{j}\!+\!\partial_{t}^2\psi_{j}\!+\!\gamma|\psi_{j}|^2\psi_{j}
\!+\!2\gamma\!\sum_{k \ne j}\!|\psi_{k}|^2\psi_{j}=i{\cal F}^{-1}(g_{j}(\omega) \hat\psi_{j})/2,  
\!\!\!
\label{Kerr1}
\end{eqnarray}        
where $\psi_{j}$ is the envelope of the electric field of the $j$th sequence, 
$1 \le j \le N$, $z$ is propagation distance, $t$ is time, $\omega$ is frequency,  
$\gamma$ is the Kerr nonlinearity coefficient, and the sum over $k$ 
extends from 1 to $N$ \cite{dimensions}. In Eq. (\ref{Kerr1}), 
$g_j(\omega)$ is the linear gain-loss experienced by the $j$th sequence, 
$\hat\psi_{j}$ is the Fourier transform of $\psi_{j}$ with respect to time, 
and ${\cal F}^{-1}$ is the inverse Fourier transform. 
The second term on the left-hand side 
of Eq. (\ref{Kerr1}) is due to second-order dispersion, 
the third term describes self-phase modulation 
and intrasequence cross-phase modulation, while the fourth term 
describes intersequence cross-phase modulation. 
The term on the right-hand side of Eq. (\ref{Kerr1}) is due to linear gain-loss. 
The optical pulses in the $j$th sequence 
are fundamental solitons of the unperturbed NLS equation 
$i\partial_z\psi_{j}+\partial_{t}^2\psi_{j}+\gamma|\psi_{j}|^2\psi_{j}=0$. 
 The envelopes of these solitons are given by 
$\psi_{sj}(t,z)=\eta_{j}\exp(i\chi_{j})\mbox{sech}(x_{j})$,
where $x_{j}=(\gamma/2)^{1/2}\eta_{j}\left(t-y_{j}-2\beta_{j} z\right)$, 
$\chi_{j}=\alpha_{j}+\beta_{j}(t-y_{j})+
\left(\gamma\eta_{j}^2/2-\beta_{j}^{2}\right)z$, 
and $\eta_{j}$, $\beta_{j}$, $y_{j}$, and $\alpha_{j}$ 
are the soliton amplitude, frequency, position, and phase.

\begin{figure}[ptb]
\begin{tabular}{cc}
\epsfxsize=10cm  \epsffile{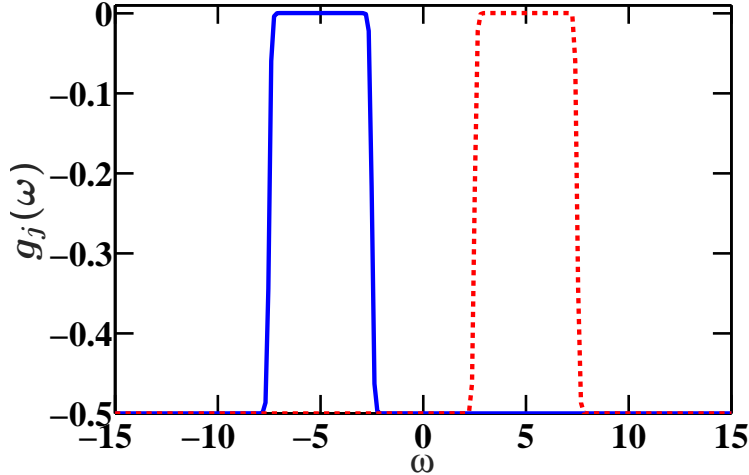} 
\end{tabular}
\caption{An example for the frequency dependent linear gain-loss 
functions $g_j(\omega)$  defined by Eq. (\ref{Kerr2}) in a two-channel waveguide coupler. 
The solid blue and dashed red lines correspond to $g_1(\omega)$ and $g_2(\omega)$, 
respectively.}        
\label{fig_add_1}
\end{figure}

Notice that Eq. (\ref{Kerr1}) describes both propagation in a single waveguide 
and propagation in a waveguide coupler, consisting of $N$ close waveguides \cite{single_waveguide}. 
In waveguide coupler transmission, each waveguide is characterized by its linear 
gain-loss function $g_j(\omega)$. The form of $g_j(\omega)$ is chosen such that 
radiation emission effects are mitigated, while the soliton patterns remain intact. 
In particular, we choose the form  
\begin{eqnarray} &&
g_j(\omega) = -g_{L} + \frac{1}{2}\left(g_{eq} + g_{L}\right)
\left[\tanh \left\lbrace \rho \left[\omega - \beta_{j}(0)+W/2\right] 
\right\rbrace 
\right.
\nonumber \\&&
\left.
- \tanh \left\lbrace \rho \left[\omega - \beta_{j}(0)- W/2\right] 
\right\rbrace\right], 
\!\!\!\!\!\!\!
\label{Kerr2}
\end{eqnarray}      
where $1 \le j \le N$, and $\beta_{j}(0)$ is the initial frequency of the $j$th sequence 
solitons. The constants $g_{L}$, $g_{eq}$, $\rho$, and $W$ satisfy $g_{L}>0$, 
$g_{eq} \ge 0$, $\rho\gg 1$, and $\Delta\beta > W > 1$, 
where $\Delta\beta$ is the intersequence frequency difference. 
We note that the condition  $\Delta\beta > 1$ is typical 
for soliton-based WDM transmission experiments 
\cite{MM98,MMN96,Nakazawa97,Nakazawa99,Nakazawa2000}.  
Figure \ref{fig_add_1} shows typical  linear gain-loss functions 
$g_{1}(\omega)$ and $g_{2}(\omega)$ for a two-channel 
waveguide coupler with $g_{L}=0.5$, $g_{eq}=3.9 \times 10^{-4}$,   
$\beta_{1}(0)=-5$, $\beta_{2}(0)=5$, $W=5$ and $\rho=10$
(these parameters are used in the numerical simulations, 
whose results are shown in Fig. \ref{fig6}).
In the limit as $\rho\gg 1$, $g_j(\omega)$ can be approximated by a step function,  
which is equal to $g_{eq}$ inside a frequency interval of width $W$ centered about 
$\beta_{j}(0)$, and to $-g_{L}$ elsewhere: 
\begin{eqnarray} &&
g_{j}(\omega) \simeq 
\left\{ \begin{array}{l l}
g_{eq} &  \mbox{ if $\beta_{j}(0)-W/2 < \omega \le \beta_{j}(0)+W/2$,}\\
g_{L} &  \mbox{elsewhere.}\\
\end{array} \right. 
\label{Kerr2A}
\end{eqnarray}     
The approximate expression (\ref{Kerr2A}) helps clarifying the advantages of using 
the linear gain-loss function (\ref{Kerr2}) for transmission stabilization. Indeed, 
the relatively strong linear loss $g_{L}$ leads to efficient suppression 
of radiative sideband generation outside of the frequency interval      
$(\beta_{j}(0)-W/2, \beta_{j}(0)+W/2]$. 
Furthermore, the relatively weak linear gain $g_{eq}$ in the frequency interval 
$(\beta_{j}(0)-W/2, \beta_{j}(0)+W/2]$ compensates for the strong loss 
outside of this interval and in this manner enables soliton propagation without amplitude decay.     
In practice, we first determine the values of $g_{L}$, $W$, 
and $\rho$ by performing simulations 
with Eqs. (\ref{Kerr1}) and (\ref{Kerr2}) with $g_{eq}=0$, 
while looking for the set that yields the longest stable propagation distance.
Once $g_{L}$, $W$, and $\rho$ are found, we determine $g_{eq}$ 
by requiring $\eta_{j}(z)=\eta_{j}(0)=\mbox{const}$ for $1 \le j \le N$ 
throughout the propagation. More specifically, we use the adiabatic perturbation 
theory for the NLS soliton (see, e.g., Ref. \cite{Iannone98}) 
to derive the following equation for the rate of change of $\eta_{j}$ with $z$ 
due to the linear gain-loss (\ref{Kerr2}):   
\begin{eqnarray} &&
\frac{{d\eta_j}}{{dz}} =
\left[ - g_L  + \left( {g_{eq} + g_L} \right) \tanh \left( {\frac{{\pi W}}
{{(8\gamma)^{1/2}\eta_j}}} \right)\right]\eta_j .
\label{Kerr2B}
\end{eqnarray}
Requiring $\eta_{j}(z)=\eta_{j}(0)=\mbox{const}$, 
we obtain the following expression for $g_{eq}$: 
\begin{eqnarray} &&
g_{eq}=
\left\lbrace
\left[\tanh \left( {\frac{{\pi W}}
{{(8\gamma)^{1/2}\eta_j(0)}}} \right)\right]^{-1}-1
\right\rbrace g_{L}.
\label{Kerr2C}
\end{eqnarray}

Since different pulse sequences propagate with different group velocities, 
the solitons undergo a large number of intersequence collisions. 
Due to the finite length of the waveguide and the finite separation  
between adjacent solitons in each sequence, the collisions are not 
completely elastic. Instead, the collisions lead to 
emission of continuous radiation with peak power 
that is inversely proportional to the intersequence frequency difference $\Delta\beta$.  
The emission of continuous radiation in multiple collisions eventually leads 
to pulse pattern distortion and to transmission destabilization. 
In the current paper, we analyze the dependence of transmission stability 
on physical parameter values and develop waveguide setups,
which lead to significant enhancement of transmission stability.  


\section{Numerical simulations}
\label{simu}

\subsection{Introduction}
\label{simu_intro}

\begin{figure}[ptb]
\begin{center}
\begin{tabular}{cc}
\epsfxsize=11.0cm  \epsffile{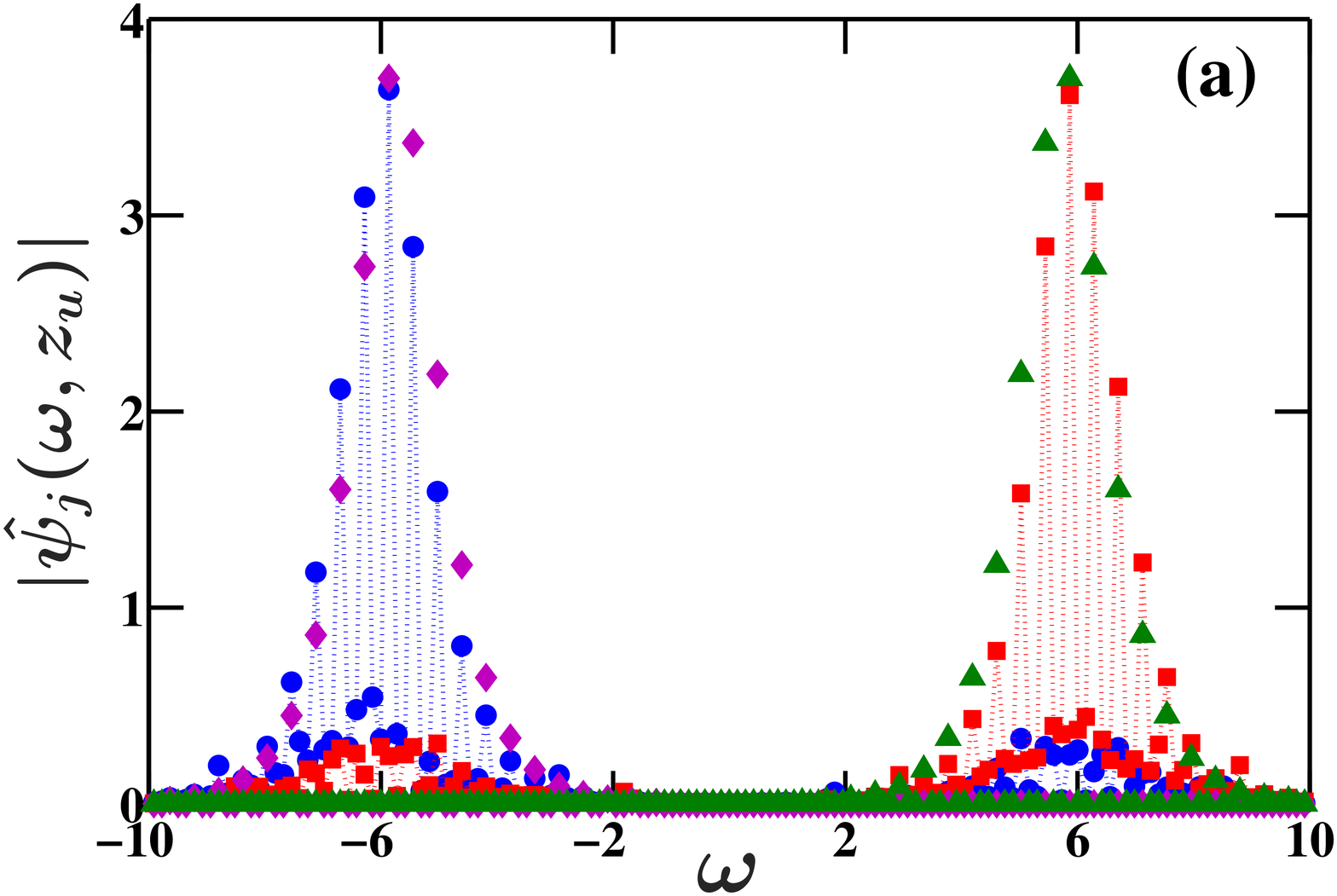} \\
\epsfxsize=11.0cm  \epsffile{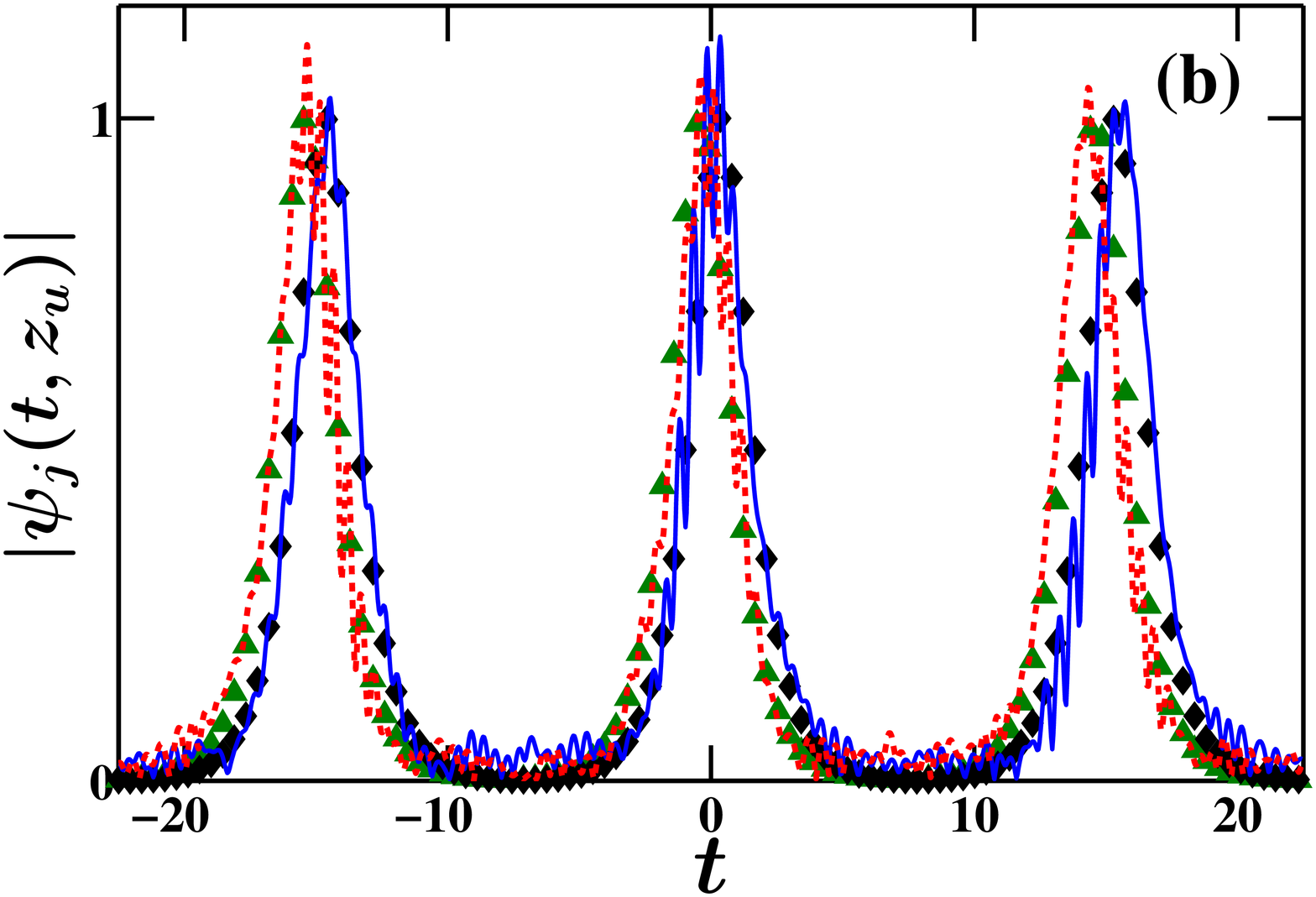} 
\end{tabular}
\end{center}
\caption{(a) The Fourier transforms of the soliton patterns at the onset 
of instability $|\hat\psi_{j}(\omega,z_{u})|$, where $z_{u}=470$, 
for two-channel transmission in a single lossless waveguide with $\gamma=2$, 
$T=15$, and $\Delta\beta=12$. The blue circles and red squares 
represent $|\hat\psi_{j}(\omega,z_{u})|$ with $j=1,2$, 
obtained by numerical solution of Eq. (\ref{Kerr1}), 
while the magenta diamonds and green triangles correspond 
to the theoretical prediction.        
(b) The soliton patterns at the onset of instability 
$|\psi_{j}(t,z_{u})|$ for two-channel transmission with the parameters used in (a). 
The solid blue and dashed red lines correspond to $|\psi_{j}(t,z_{u})|$ with $j=1,2$, 
obtained by the simulations, while the black diamonds and green triangles 
correspond to the theoretical prediction.}
\label{fig1}
\end{figure}

To investigate transmission stability, we numerically integrate the system (\ref{Kerr1}),  
using the split-step method with periodic boundary conditions \cite{Agrawal2001}. 
The use of periodic boundary conditions means that  
the simulations describe pulse dynamics in a closed waveguide loop. 
The initial condition is in the form of $N$ periodic sequences 
of $2K+1$ solitons with amplitudes $\eta_{j}(0)$, 
frequencies $\beta_{j}(0)$, and zero phases:  
\begin{eqnarray} &&
\psi_{j}(t,0)\!=\!\sum_{k=-K}^{K}
\frac{\eta_{j}(0)\exp[i\beta_{j}(0)(t-kT)]}
{\cosh[(\gamma/2)^{1/2}\eta_{j}(0)(t-kT)]},
\label{Kerr3}
\end{eqnarray}
where $1\le j \le N$, $T$ is the time-slot width, and $N=2$ or $N=3$. 
This initial condition represents the typical situation in multichannel 
soliton-based transmission \cite{Agrawal2001,Mollenauer2006,Iannone98}. 
To maximize the stable propagation distance, we choose 
$\beta_{1}(0)$=-$\beta_{2}(0)$ for a two-channel system, 
and $\beta_{1}(0)$=-$\beta_{3}(0)$ and $\beta_{2}(0)=0$ 
for a three-channel system.  
This choice is based on extensive numerical simulations with Eq. (\ref{Kerr1}) 
with the right-hand-side set equal to zero and different values of $\beta_j(0)$. 
For concreteness, we present here 
the results of numerical simulations with parameter values 
$T=15$, $\eta_{j}(0)=1$, $K=1$, and a final transmission distance $z_{f}=5000$. 
We emphasize, however, that similar results are obtained with 
other values of the physical parameters. 
That is, the results reported in this section are not very sensitive 
to the values of $K$, $\eta_{j}(0)$, and $T$, as long as $\eta_{j}(0)$ 
is not much smaller than or much larger than 1, and as long as $T>10$.

\subsection{Two-channel transmission}
\label{2_channel}

\begin{figure}[ptb]
\begin{center}
\begin{tabular}{cc}
\epsfxsize=11.5cm  \epsffile{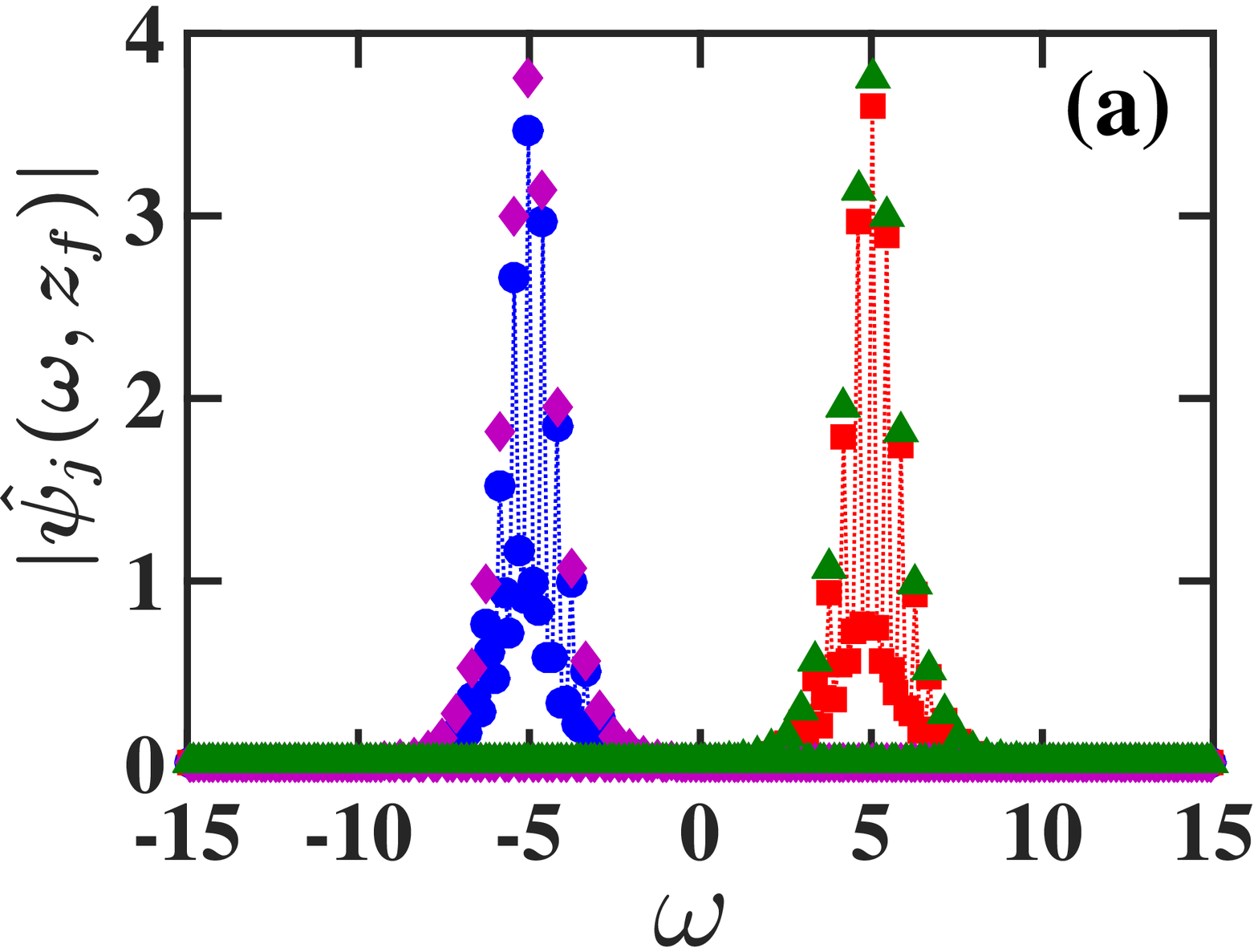} \\
\epsfxsize=11.5cm  \epsffile{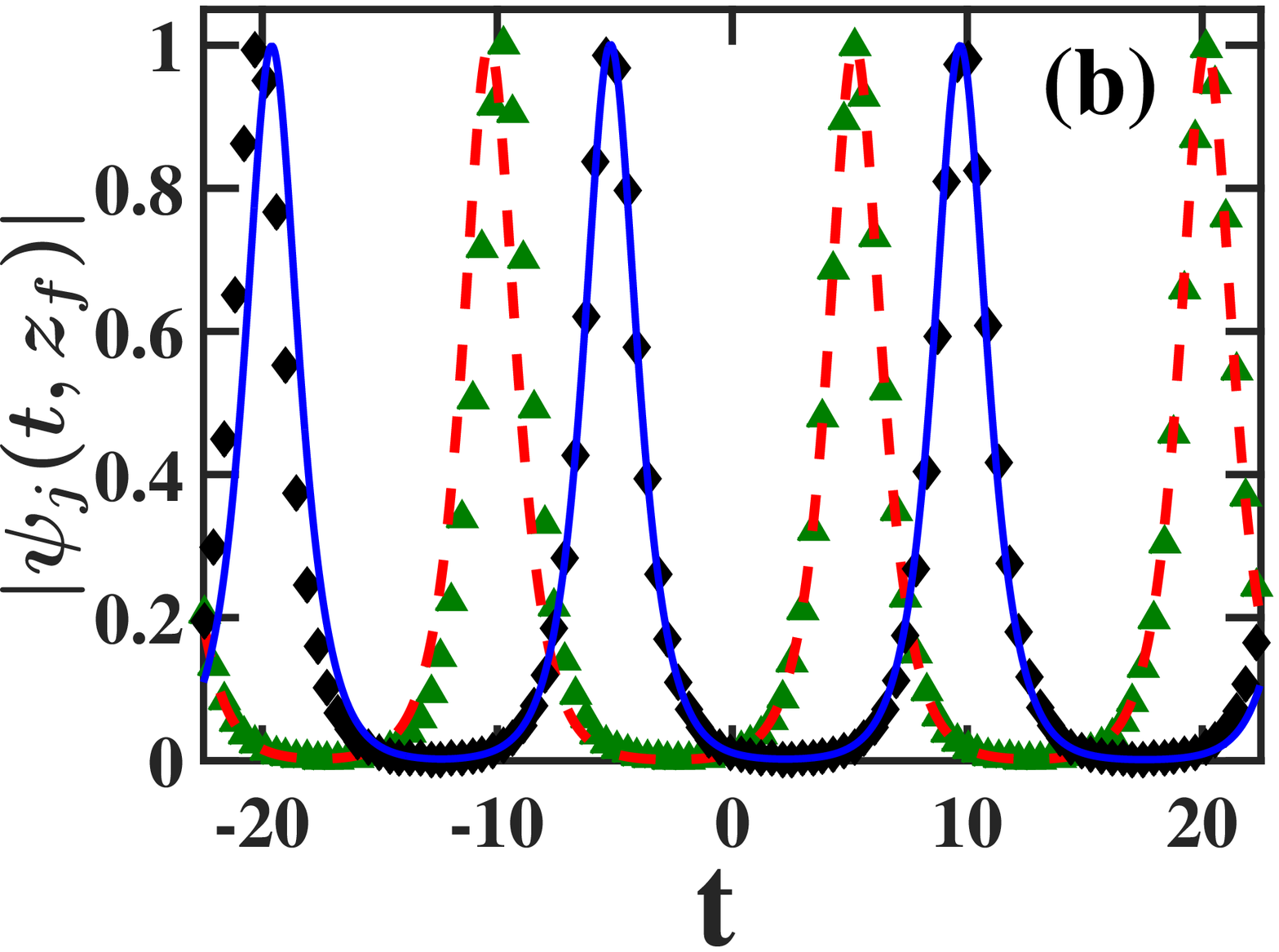} 
\end{tabular}
\end{center}
\caption{(a) The Fourier transforms of the soliton patterns at the final propagation 
distance $|\hat\psi_{j}(\omega,z_{f})|$, where $z_{f}=20000$, 
in the case where each soliton sequence propagates on its own 
through a lossless waveguide. 
The values of the physical parameters are $\beta_{1}(0)=-5$,  
$\beta_{2}(0)=5$, $\gamma=2$, and $T=15$. 
The symbols are the same as in Fig. \ref{fig1}(a). 
(b) The soliton patterns at the final propagation distance $|\psi_{j}(t,z_{f})|$ 
for the single-sequence propagation setup considered in (a). 
The symbols are the same as in Fig. \ref{fig1}(b).}
\label{fig_add_2}
\end{figure}    
    
We start by considering two-channel transmission in a single lossless waveguide. 
Simulations with Eq. (\ref{Kerr1}) with $N=2$  
show stable propagation at short-to-intermediate distances and 
transmission destabilization at long distances. As seen in Fig. \ref{fig1}, 
the instability first appears as fast temporal oscillations in the soliton patterns, 
which is caused by resonant formation of radiative sidebands 
with frequencies $\beta_{2}(0)$ for $j=1$ and $\beta_{1}(0)$ for $j=2$. 
The growth of the radiative sidebands with increasing $z$ eventually leads  
to the destruction of the soliton patterns. 
We note that when each soliton sequence propagates through the waveguide 
on its own, no radiative sidebands develop and no instability is observed 
up to distances as large as $z=20000$ (see Fig. \ref{fig_add_2}). 
Thus, the instability is caused by the Kerr-induced interaction in interchannel collisions, 
i.e., it is associated with the intersequence cross-phase modulation terms 
$2\gamma|\psi_{k}|^2\psi_{j}$ in Eq. (\ref{Kerr1}).

\begin{figure}[ptb]
\begin{tabular}{cc}
\epsfxsize=12cm  \epsffile{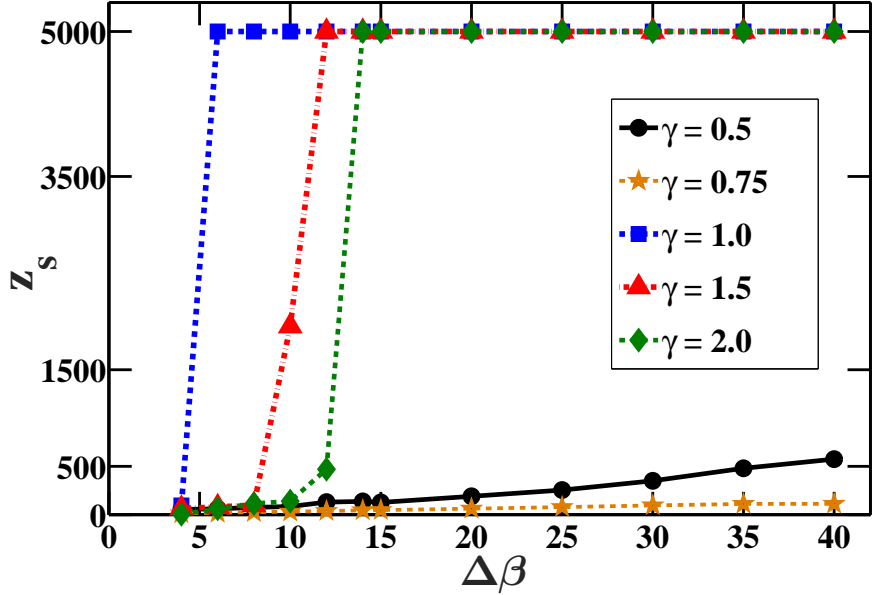} 
\end{tabular}
\caption{Stable propagation distance $z_{s}$ vs frequency spacing 
$\Delta\beta$ for two-channel transmission in a single lossless waveguide with $T=15$.
The black circles, orange stars, blue squares, red triangles, 
and green diamonds represent the results obtained 
by the simulations for $\gamma=0.5, 0.75, 1.0, 1.5$, and 2.0, respectively.}        
\label{fig2}
\end{figure}           

An important question about the transmission concerns the dependence 
of transmission stability on the value of the Kerr nonlinearity 
coefficient. In particular, we would like to find if there is 
an optimal value of $\gamma$, which leads to minimization of radiative 
sideband generation and to maximization of transmission stability.      
To answer this question, we define the stable propagation distance $z_{s}$  
as the distance $z_{u}$ at which instability develops, if $z_{u}<z_{f}$, 
and as $z_{f}$, if no instability is observed throughout the propagation.     
That is, $z_s=z_u$, if $z_u < z_f$ (instability is observed), and $z_s=z_f$, 
if $z_u \ge z_f$ (instability is not observed). 
We then carry out simulations with Eq.  (\ref{Kerr1}) 
for $N=2$, $0.5 \le \gamma \le 2$, and $4 \le \Delta\beta \le 40$, and plot 
$z_{s}$ vs frequency spacing $\Delta\beta$. 
The results of the simulations are shown in Fig. \ref{fig2}. 
It is seen that $z_{s}$ increases with increasing $\Delta\beta$, 
in accordance with the decrease of intersequence cross-phase modulation effects 
with increasing frequency spacing \cite{MM98}. 
Moreover, for all frequency differences $\Delta\beta$ in the interval $4 \le \Delta\beta \le 40$, 
the $z_{s}$ values obtained with $\gamma=1$ are larger than or equal to the $z_{s}$ 
values achieved with $\gamma \ne 1$. Thus, $\gamma=1$ is the optimal value 
of the Kerr nonlinearity coefficient. 
Based on these results and results of simulations with other 
sets of physical parameters, we conclude that for two-channel systems, 
there indeed exists an optimal value of the Kerr nonlinearity coefficient, 
which minimizes radiative sideband generation and yields the 
longest stable propagation distance.

\begin{figure}[ptb]
\begin{center}
\begin{tabular}{cc}
\epsfxsize=11.0cm  \epsffile{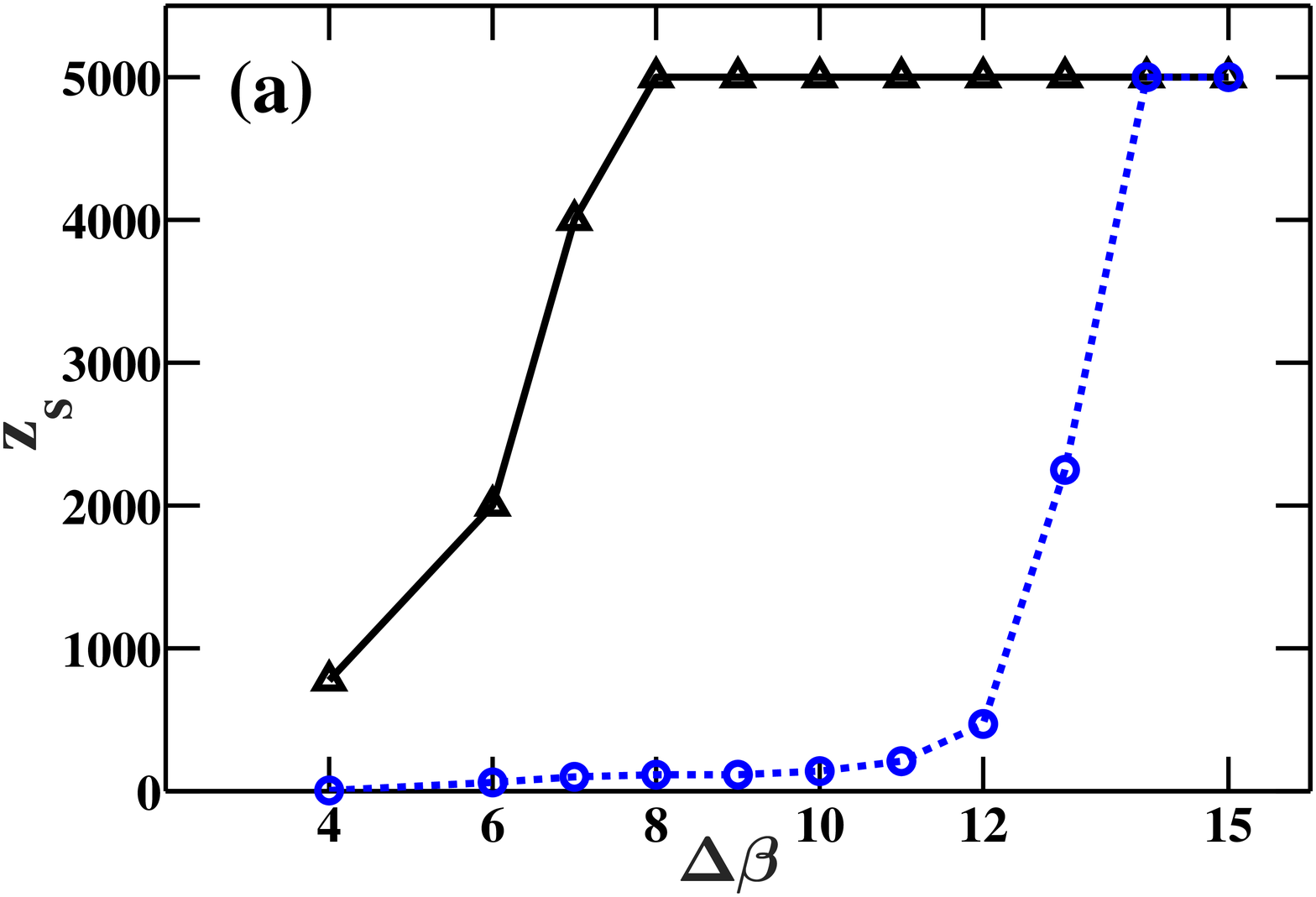}\\
\epsfxsize=11.0cm  \epsffile{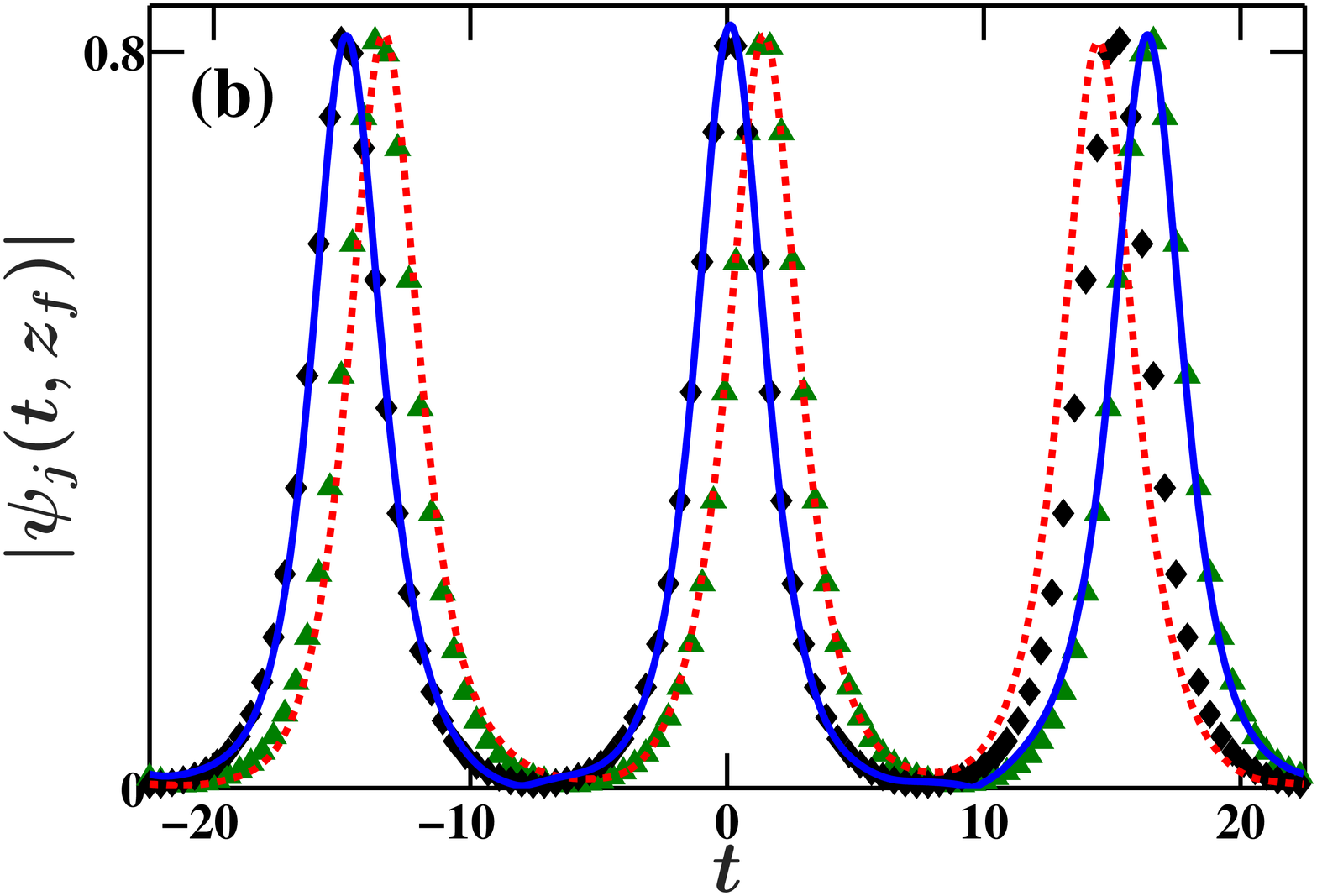}
\end{tabular}
\end{center}
\caption{(a) Stable propagation distance $z_{s}$ vs frequency 
spacing $\Delta\beta$ for two-channel waveguide coupler transmission with 
frequency dependent linear loss and $\gamma=2$, $T=15$, 
$g_{L}=0.5$, $g_{eq}=0$, $\rho=10$, and $W=\Delta\beta/2$.   
The solid black line is the result obtained by 
numerical solution of Eqs. (\ref{Kerr1}) and (\ref{Kerr2}).   
The dashed blue line is the result obtained by the 
simulations for  two-channel transmission in a single lossless 
waveguide with $\gamma=2$ and $T=15$.       
(b) The final pulse patterns $|\psi_{j}(t,z_{f})|$, 
where $z_{f}=5000$, in two-channel waveguide coupler transmission 
with $\Delta\beta=12$. The symbols are the same as in Fig. \ref{fig1}(b).}        
\label{fig3}
\end{figure}

Since the radiative sideband for the $j$th sequence forms at frequency 
$\beta_{k}(0)$ of the other sequence, it is very difficult to suppress radiative 
instability in a single waveguide by frequency dependent gain-loss. 
The situation is very different in waveguide coupler transmission, since in this case 
one can employ a different gain-loss profile for each waveguide, with strong loss 
for all frequencies outside of a frequency interval centered about $\beta_{j}(0)$. 
We therefore turn to consider waveguide couplers with frequency dependent 
linear loss, and show that in this case, significant enhancement of transmission stability
can be achieved, even for non-optimal $\gamma$ values.  
For this purpose, we numerically solve 
Eqs. (\ref{Kerr1}) and (\ref{Kerr2}) with $N=2$ and $g_{eq}=0$ for different 
$\gamma$ values and $4 \le \Delta\beta \le 15$.   
Here we present the results obtained for $\gamma=2$,  
$T=15$, $g_{L}=0.5$, $\rho=10$, and $W=\Delta\beta/2$. 
Similar results are obtained with other choices of the physical parameter values. 
Figure \ref{fig3}(a) shows the stable propagation distance $z_{s}$ 
vs frequency spacing $\Delta\beta$ as obtained in the simulations 
for two-channel waveguide coupler transmission along with the value 
obtained for transmission in a single lossless waveguide. 
We note that for $\Delta\beta \ge 8$, 
$z_{s}=z_{f}=5000$, i.e., the transmission is stable throughout 
the propagation. Moreover, the $z_{s}$ values obtained 
for waveguide coupler transmission are larger than the values 
obtained for single waveguide transmission  by factors 
ranging between 172.2 for $\Delta\beta=4$ and 
2.22 for $\Delta\beta=13$. Additionally, as seen in  
Fig. \ref{fig3}(b), the solitons retain their shape 
throughout the propagation.

\begin{figure}[ptb]
\begin{center}
\begin{tabular}{cc}
\epsfxsize=7.5cm  \epsffile{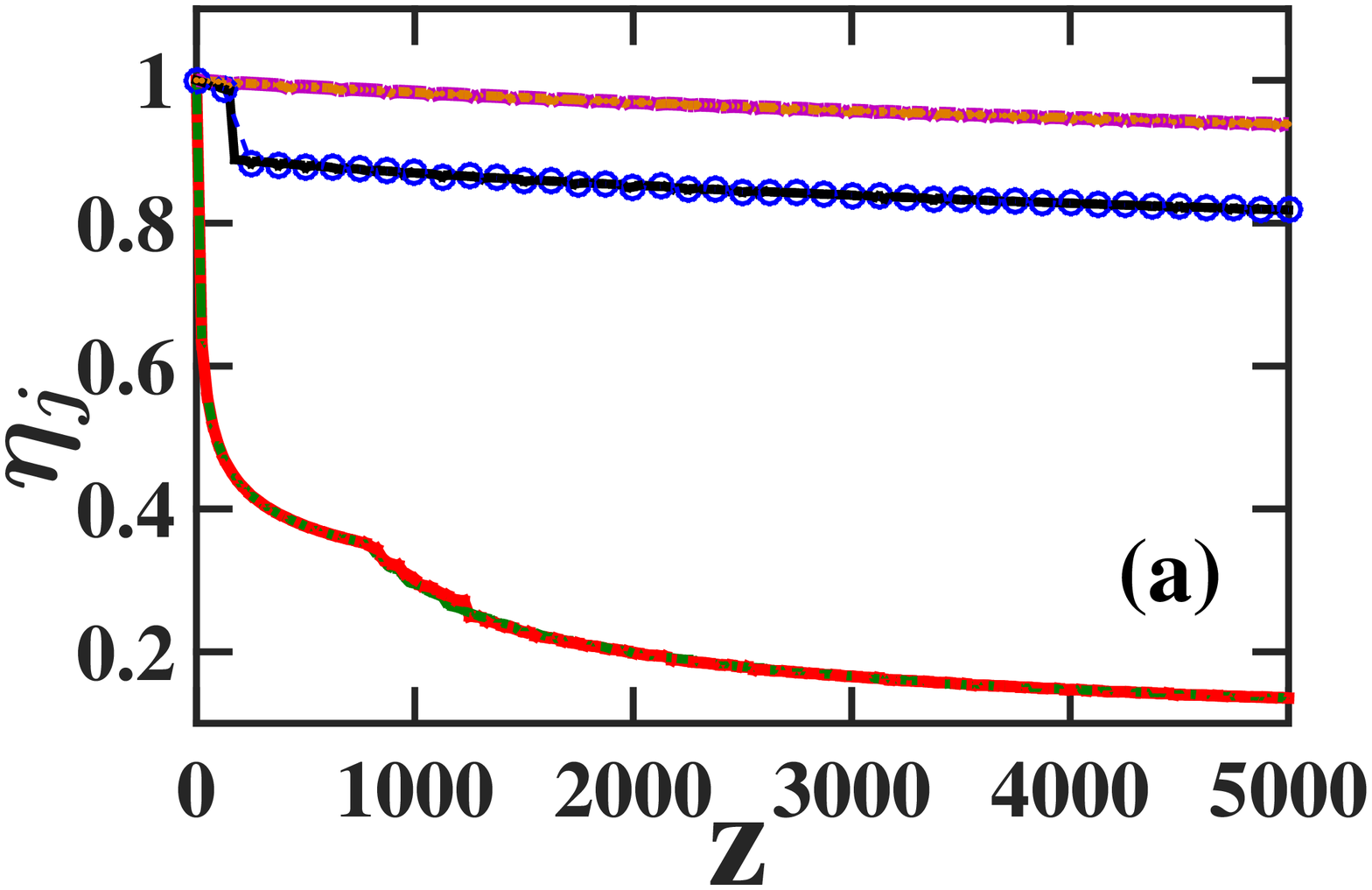} \\
\epsfxsize=7.1cm  \epsffile{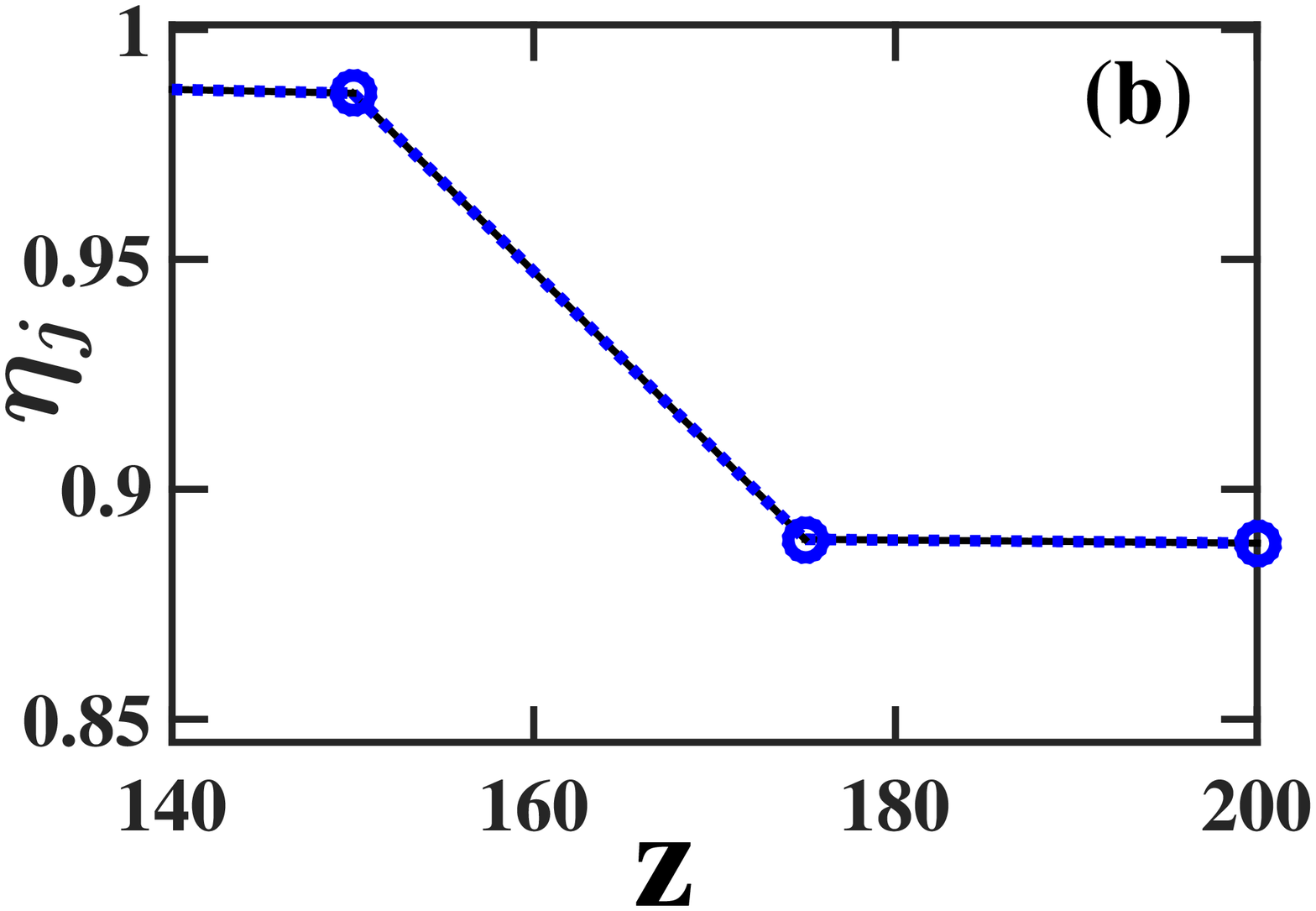} \\
\epsfxsize=7.1cm  \epsffile{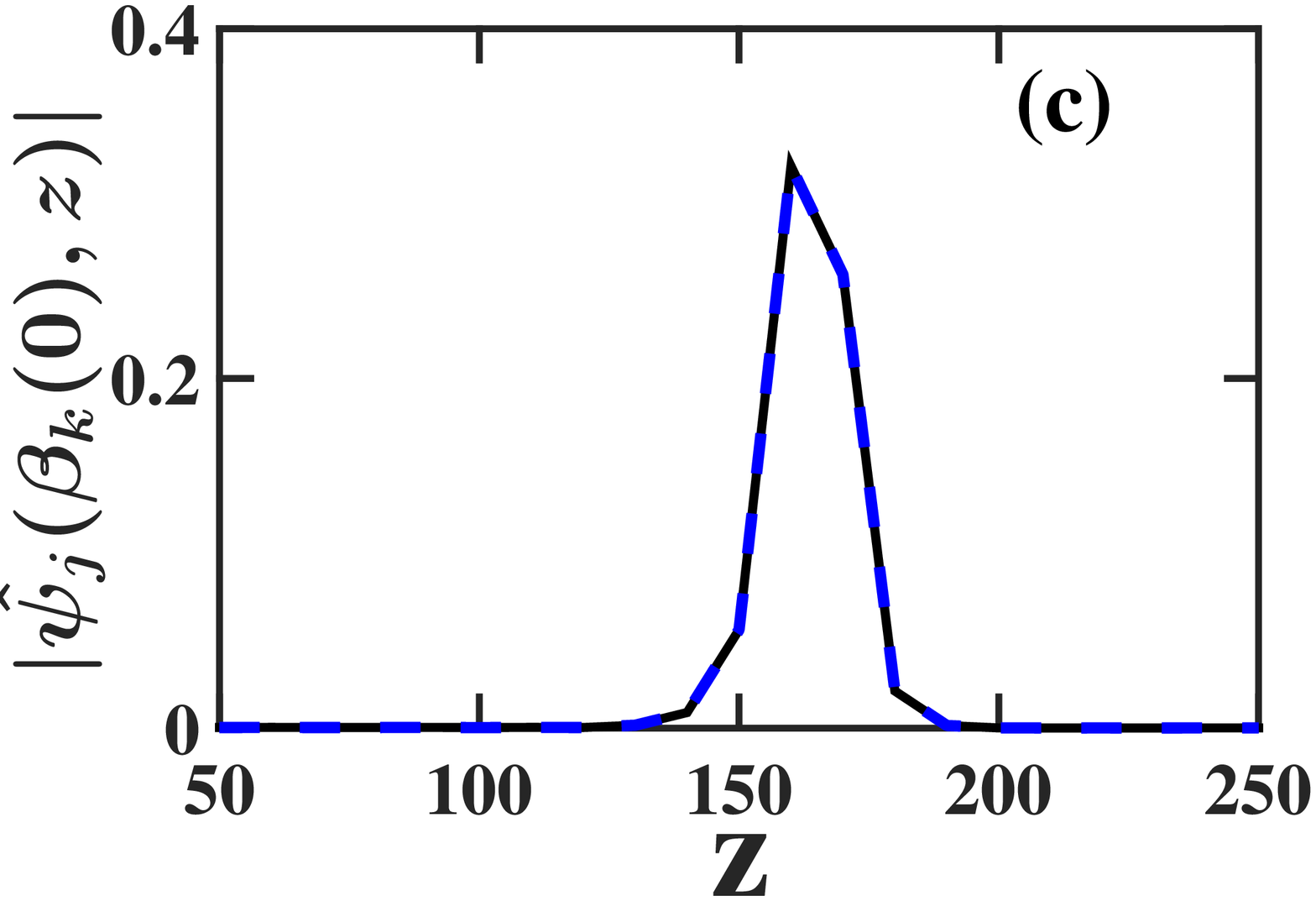}
\end{tabular}
\end{center}
\caption{ (a) The $z$ dependence of soliton amplitudes $\eta_{j}$ 
for two-channel waveguide coupler transmission with frequency dependent 
linear loss and $\gamma=2$, $T=15$, $g_{L}=0.5$, $g_{eq}=0$, 
$\rho=10$, and $W=\Delta\beta/2$.  
The solid red, solid black, and dashed-dotted purple curves correspond 
to $\eta_{1}(z)$ obtained by numerical simulations with 
Eqs. (\ref{Kerr1}) and (\ref{Kerr2}) for $\Delta\beta=4$, 
$\Delta\beta=12$, and $\Delta\beta=14$. 
The dashed-dotted-dotted green, circle-dashed blue, and short 
dashed-dotted orange curves represent $\eta_{2}(z)$ obtained 
by the simulations for $\Delta\beta=4$, $\Delta\beta=12$, 
and $\Delta\beta=14$. 
(b) Magnified versions of the $\eta_{j}(z)$ 
curves for $\Delta\beta=12$ in the interval $140 \le z \le 200$. 
(c) The $z$ dependence of radiative sideband amplitudes 
$|\hat\psi_{1}(\beta_{2}(0),z)|$ (solid black line)
and $|\hat\psi_{2}(\beta_{1}(0),z)|$ (dashed blue line), 
obtained by the simulations for $\Delta\beta=12$.}        
\label{fig4}
\end{figure}

\begin{figure}[ptb]
\begin{center}
\begin{tabular}{cc}
\epsfxsize=11cm  \epsffile{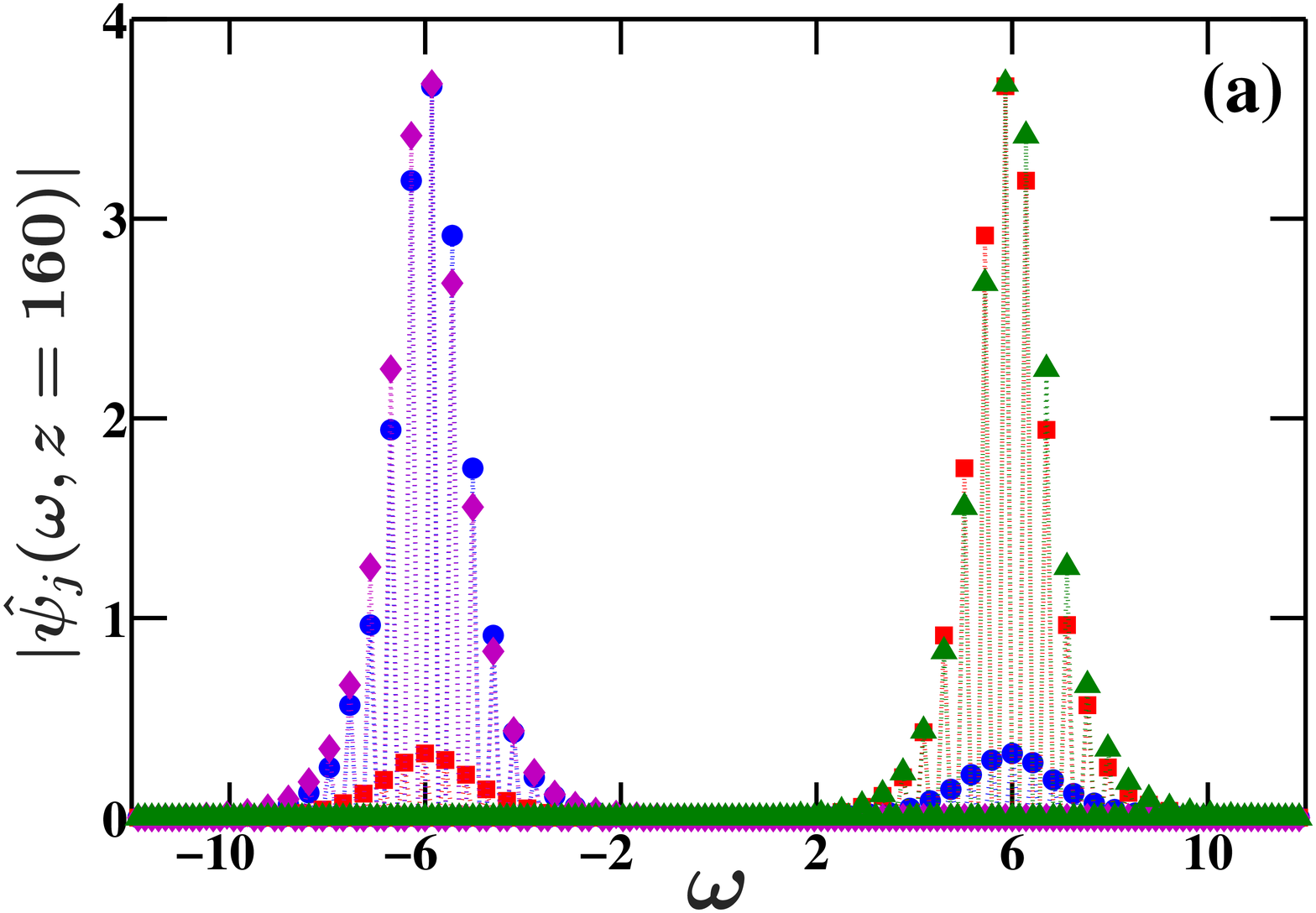} \\
\epsfxsize=11cm  \epsffile{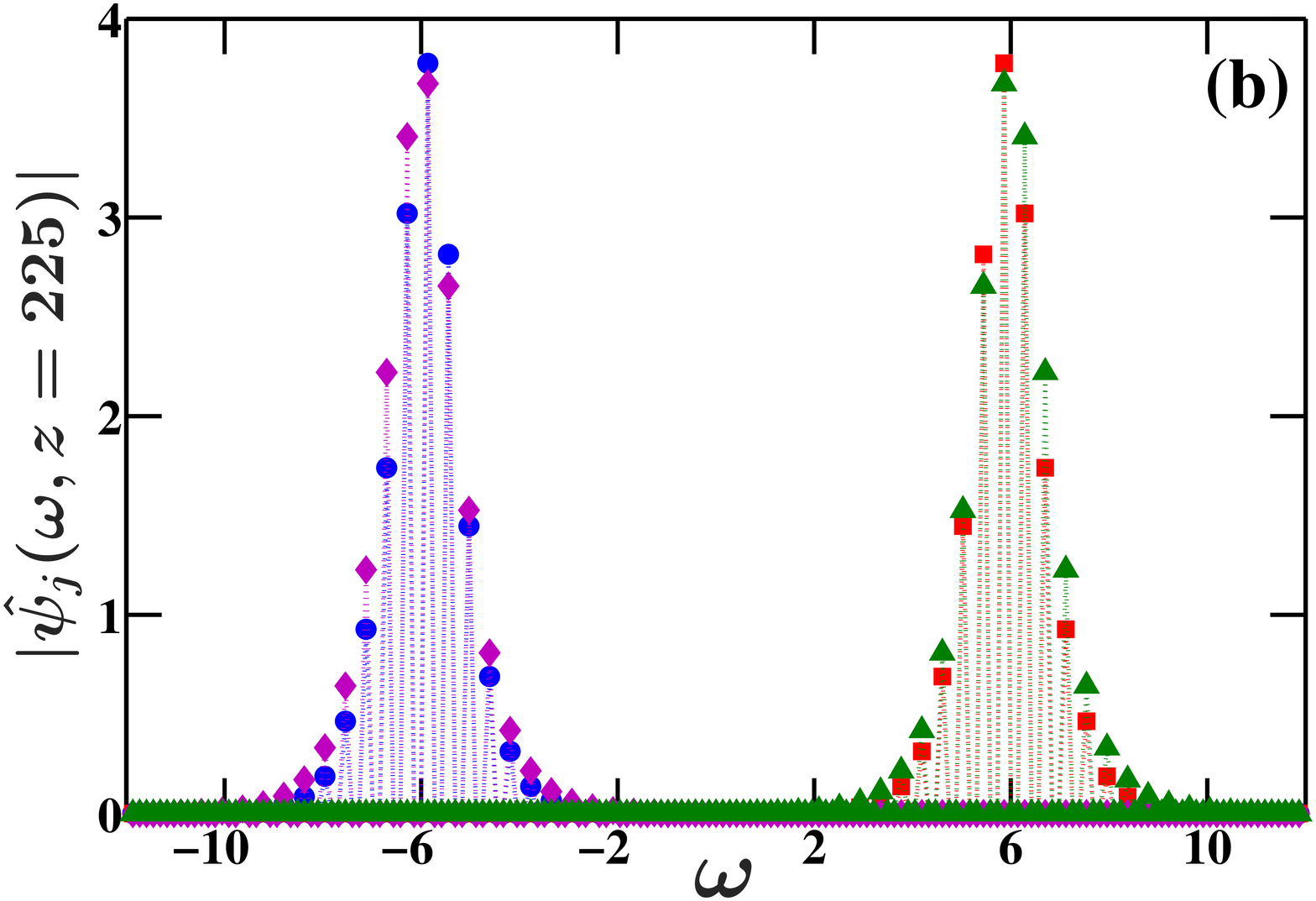} 
\end{tabular}
\end{center}
\caption{The Fourier transforms of the soliton patterns $|\hat\psi_{j}(\omega,z)|$ 
in two-channel waveguide coupler transmission with frequency dependent linear loss 
for $\Delta\beta=12$ and the same values of $\gamma$, $T$, $g_{L}$, 
$g_{eq}$, $\rho$, and $W$ as in Fig. \ref{fig4}. 
(a) $|\hat\psi_{j}(\omega,z)|$ at $z=160$. 
(b) $|\hat\psi_{j}(\omega,z)|$ at $z=225$. 
The symbols are the same as in Fig. \ref{fig1}(a)}        
\label{fig5}
\end{figure}

We now turn to analyze the $z$ dependence of soliton amplitudes 
for propagation in the waveguide coupler, since this analysis 
provides insight into the processes involved in transmission stabilization.  
We find three remarkably different dependences of soliton 
amplitudes on $z$ in the frequency spacing intervals $4 \le \Delta\beta <8$, 
$8 \le \Delta\beta <14$, and $\Delta\beta \ge 14$.    
Figure \ref{fig4}(a) shows the $\eta_{j}(z)$ 
curves obtained by the simulations for three representative cases, 
$\Delta\beta=4$, $\Delta\beta=12$, and $\Delta\beta=14$. 
For $\Delta\beta=4$ and $\Delta\beta=14$,  
the soliton amplitudes decrease gradually to their final values.  
In contrast, for $\Delta\beta=12$, soliton amplitudes 
gradually decrease for $0 \le z < 150$, but then undergo 
a steep decrease in the interval $150 \le z \le 175$, followed 
by another gradual decrease for $175 < z  \le 5000$
[see Figures \ref{fig4}(a) and \ref{fig4}(b)]. 
To explain the abrupt decrease of $\eta_{j}(z)$ in the interval 
$150 \le z \le 175$, we analyze the $z$ dependence of radiative sideband amplitudes, 
defined as $|\hat\psi_{1}(\beta_{2}(0),z)|$ 
and $|\hat\psi_{2}(\beta_{1}(0),z)|$ for $j=1$ and $j=2$, respectively. 
As seen in Figures \ref{fig4}(c) and \ref{fig5}, 
sideband amplitudes exhibit different behavior for $0 \le z < 120$, $120 \le z < 200$, 
and $200 \le z \le 5000$, which correspond to the three intervals observed 
for soliton amplitude dynamics. More specifically, for $0 \le z < 120$, 
sideband amplitudes are smaller than $10^{-3}$ and are slowly increasing,         
for $120 \le z < 200$, sideband amplitudes increase up to a maximum of $0.321$ 
at $z=160$ and then decrease to below $10^{-4}$ at $z=200$, 
while for $200 \le z \le 5000$, sideband amplitudes remain smaller than $2\times 10^{-4}$.  
Thus, the steep drop of $\eta_{j}(z)$ for $150 \le z \le 175$ 
is related to the growth and subsequent decay of the radiative sidebands 
in the interval $120 \le z < 200$. This can be explained by noting that 
as the radiative sidebands grow, energy is rapidly transferred from a localized soliton 
form to a nonlocalized form, which is accompanied by the steep decay of soliton amplitudes. 
Additionally, the fast decay of the sidebands is a result of the strong linear loss 
$g_{L}$ at frequencies $\beta_{2}(0)$ for $j=1$ and $\beta_{1}(0)$ for $j=2$.    
Note that the sharp drop in $\eta_{j}(z)$ and the associated growth and disappearance 
of the radiative sidebands can be described as a dynamic phase transition, 
which is similar to the transition of one phase of matter to another. 
Indeed, one can consider the solitons and the radiation to be two 
different ``phases''. The abrupt disappearance of the radiation 
due to the presence of linear loss 
can then be viewed as a transition from an unstable transmission state, 
in which both phases exist in the waveguide, to a stable state, 
in which only the soliton ``phase'' exists.

\begin{figure}[ptb]
\begin{center}
\begin{tabular}{cc}
\epsfxsize=7.6cm  \epsffile{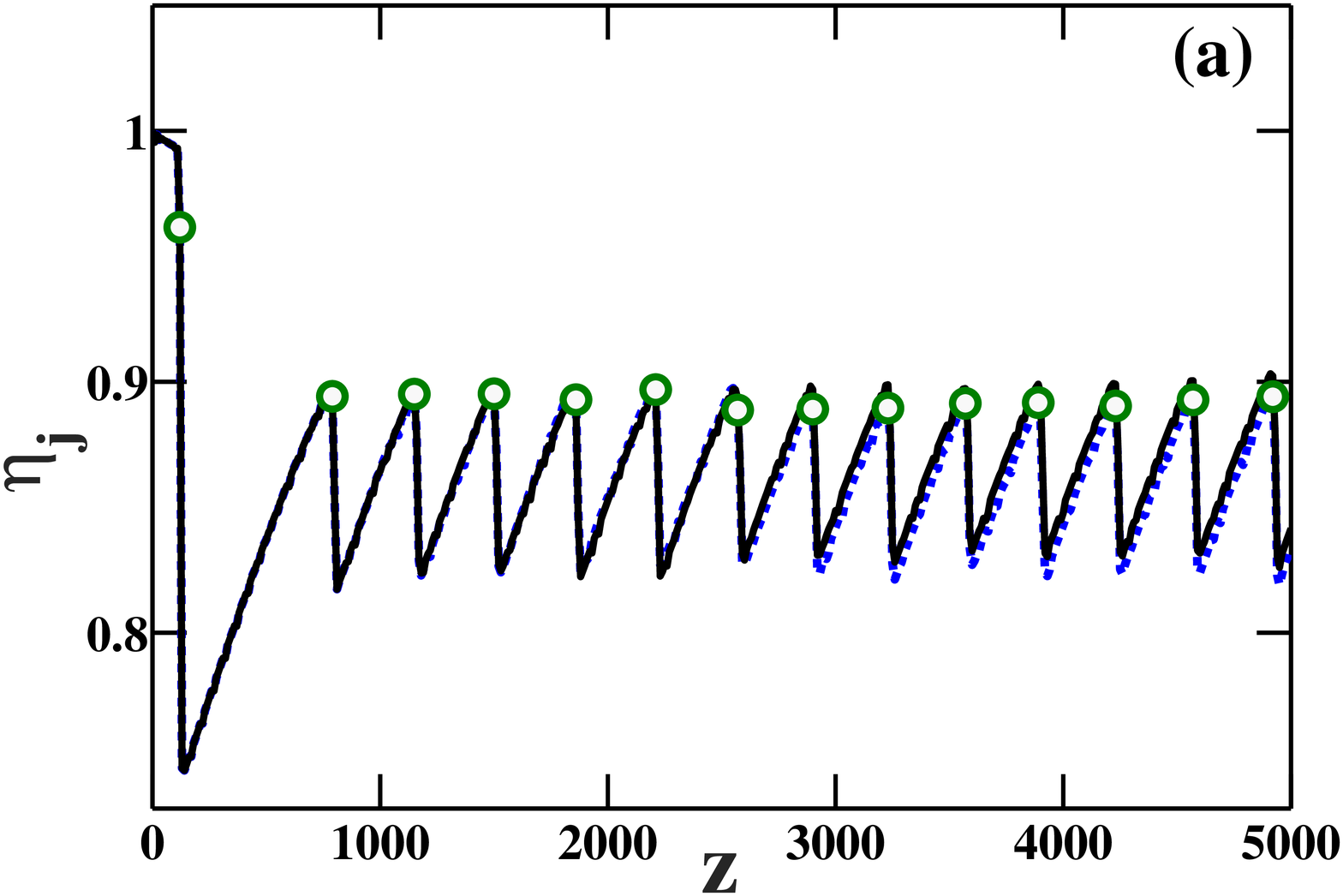}\\
\epsfxsize=7.6cm  \epsffile{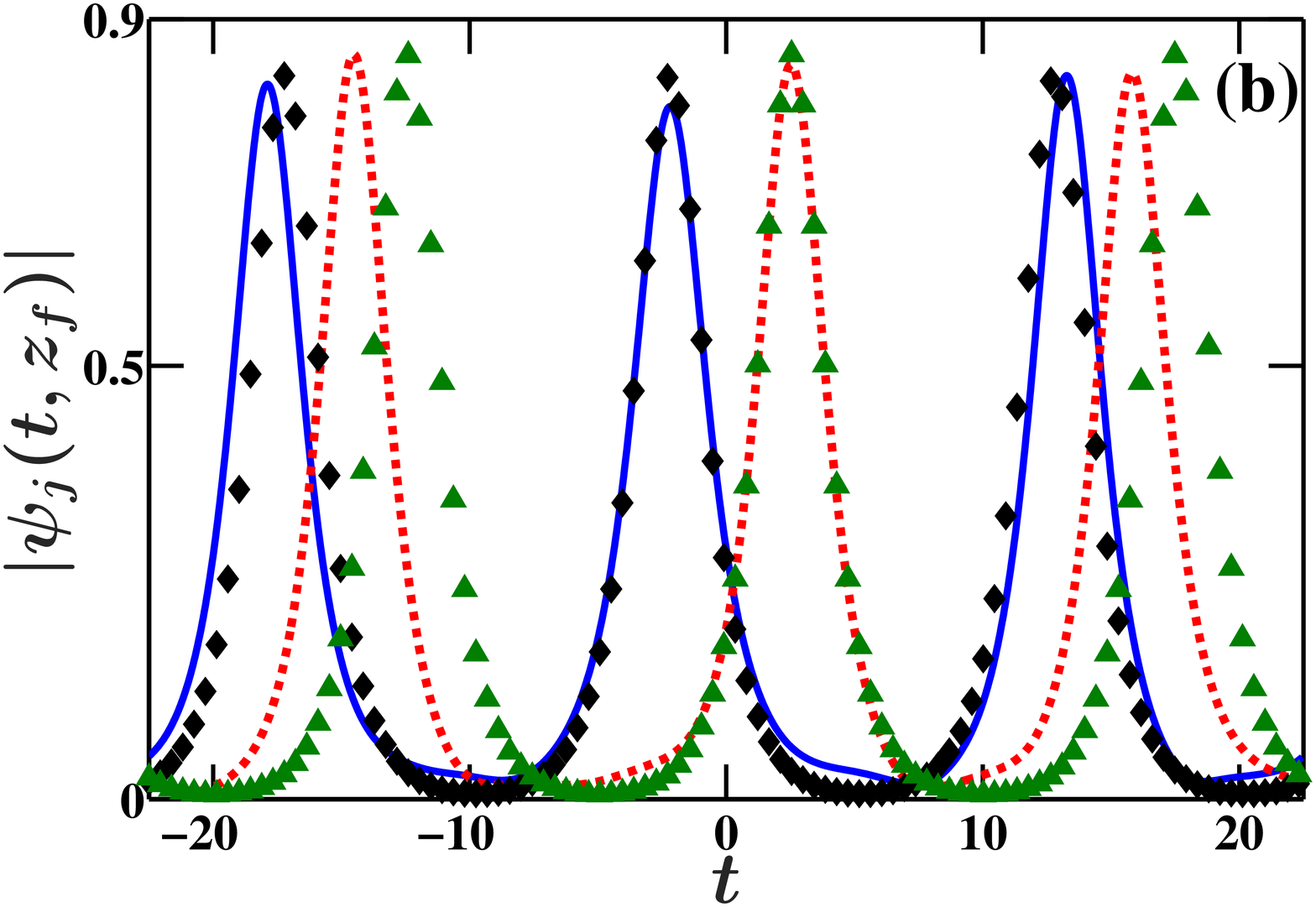}\\
\epsfxsize=7.6cm  \epsffile{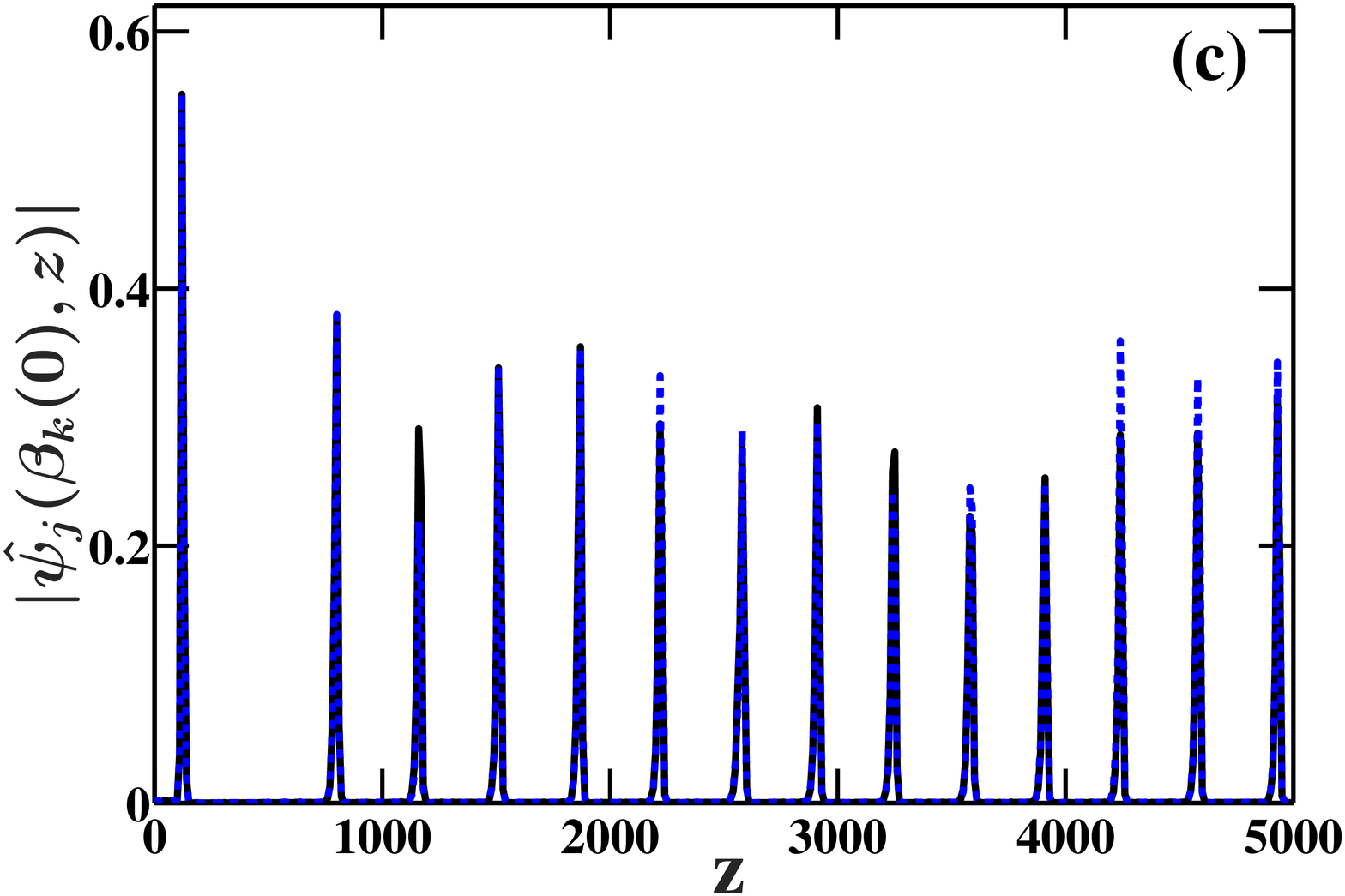}
\end{tabular}
\end{center}
\caption{(a) The $z$ dependence of soliton amplitudes $\eta_{j}$ 
for two-channel waveguide coupler transmission with frequency dependent linear 
gain-loss and $\gamma=2$, $T=15$, $\Delta\beta=10$,   
$g_{L}=0.5$, $g_{eq}=3.9 \times 10^{-4}$, $\rho=10$, and $W=5$. 
The solid black and dashed blue lines correspond to $\eta_{j}(z)$ with $j=1,2$, 
as obtained by numerical solution of Eqs. (\ref{Kerr1}) and (\ref{Kerr2}). 
The green circles indicate the distances at which 
radiative sideband amplitudes attain their maxima. 
(b) The final pulse patterns $|\psi_{j}(t,z_{f})|$, where $z_{f}=5000$.    
The symbols are the same as in Fig. \ref{fig1}(b). 
(c) The $z$ dependence of radiative sideband amplitudes 
$|\hat\psi_{1}(\beta_{2}(0),z)|$ (solid black line)
and $|\hat\psi_{2}(\beta_{1}(0),z)|$ (dashed blue line), obtained by the simulations.}        
\label{fig6}
\end{figure}

The waveguide couplers with net linear loss
have a major disadvantage due to the decay of soliton amplitudes. 
This problem can be overcome in waveguide couplers with linear gain-loss 
by introducing the net linear gain $g_{eq}$ at a frequency interval of width $W$  
centered about $\beta_{j}(0)$. We investigate two-channel soliton transmission 
in waveguide couplers with linear gain-loss by performing simulations with 
Eqs. (\ref{Kerr1}) and (\ref{Kerr2}) with $N=2$ and $g_{eq}>0$. 
To enable comparison with the results of Figures \ref{fig3} and \ref{fig4}, 
we discuss the results of simulations with the same parameter values, 
i.e., $\gamma=2$, $T=15$, $g_{L}=0.5$, $\rho=10$, and $W=\Delta\beta/2$. 
We find that soliton amplitudes exhibit different dynamic behavior 
in the frequency spacing intervals $4 \le \Delta\beta <8$, 
$8 \le \Delta\beta <14$, and $\Delta\beta \ge 14$, 
which are the same intervals observed for waveguide couplers with net linear loss. 
For $4 \le \Delta\beta <8$, amplitude values are approximately constant 
until transmission destabilization, while for $\Delta\beta \ge 14$, 
the amplitudes are approximately constant throughout the propagation. 
In contrast, for $8 \le \Delta\beta <14$, the amplitudes exhibit stable 
oscillations throughout the propagation. Figure \ref{fig6}(a) 
shows the oscillatory dynamics for $\Delta\beta=10$ and $g_{eq}=3.9 \times 10^{-4}$. 
As can be seen, the amplitudes undergo a steep decrease, 
followed by oscillations about the value $\eta_{s}=0.86$. 
Additionally, as seen in Fig. \ref{fig6}(b), pulse distortion 
at $z_{f}=5000$ is small, although the solitons within each sequence 
experience position shifts relative to one another. 
To check if the oscillations of soliton amplitudes are caused by radiative sideband dynamics,  
we analyze the $z$ dependence of radiative sideband amplitudes 
$|\hat\psi_{1}(\beta_{2}(0),z)|$ and $|\hat\psi_{2}(\beta_{1}(0),z)|$. 
As seen in Fig. \ref{fig6}(c), the amplitudes of the radiative sidebands experience alternating 
``periods'' of growth and decay. Furthermore, the points where the sidebands are maximal 
are located near the beginnings of the relatively short intervals, where soliton amplitudes 
are decreasing [see Fig. \ref{fig6}(a)]. Based on these observations, 
we conclude that the oscillatory dynamics of soliton amplitudes is 
caused by decay and regeneration of the radiative sidebands. 
This can be explained by noting that as the sidebands grow, 
energy is transferred from a localized soliton form to a nonlocalized form. 
The strong linear loss $g_{L}$ outside the central frequency intervals 
leads to relatively fast decay of the radiative sidebands, 
which is accompanied by a decrease in soliton amplitudes. 
Furthermore, the weak linear gain $g_{eq}$ at the central frequency intervals 
leads to slow growth of soliton amplitudes at the subsequent waveguide spans
and to the observed oscillatory dynamics.

\subsection{Three-channel transmission}
\label{3_channel}
\begin{figure}[ptb]
\begin{center}
\begin{tabular}{cc}
\epsfxsize=11cm  \epsffile{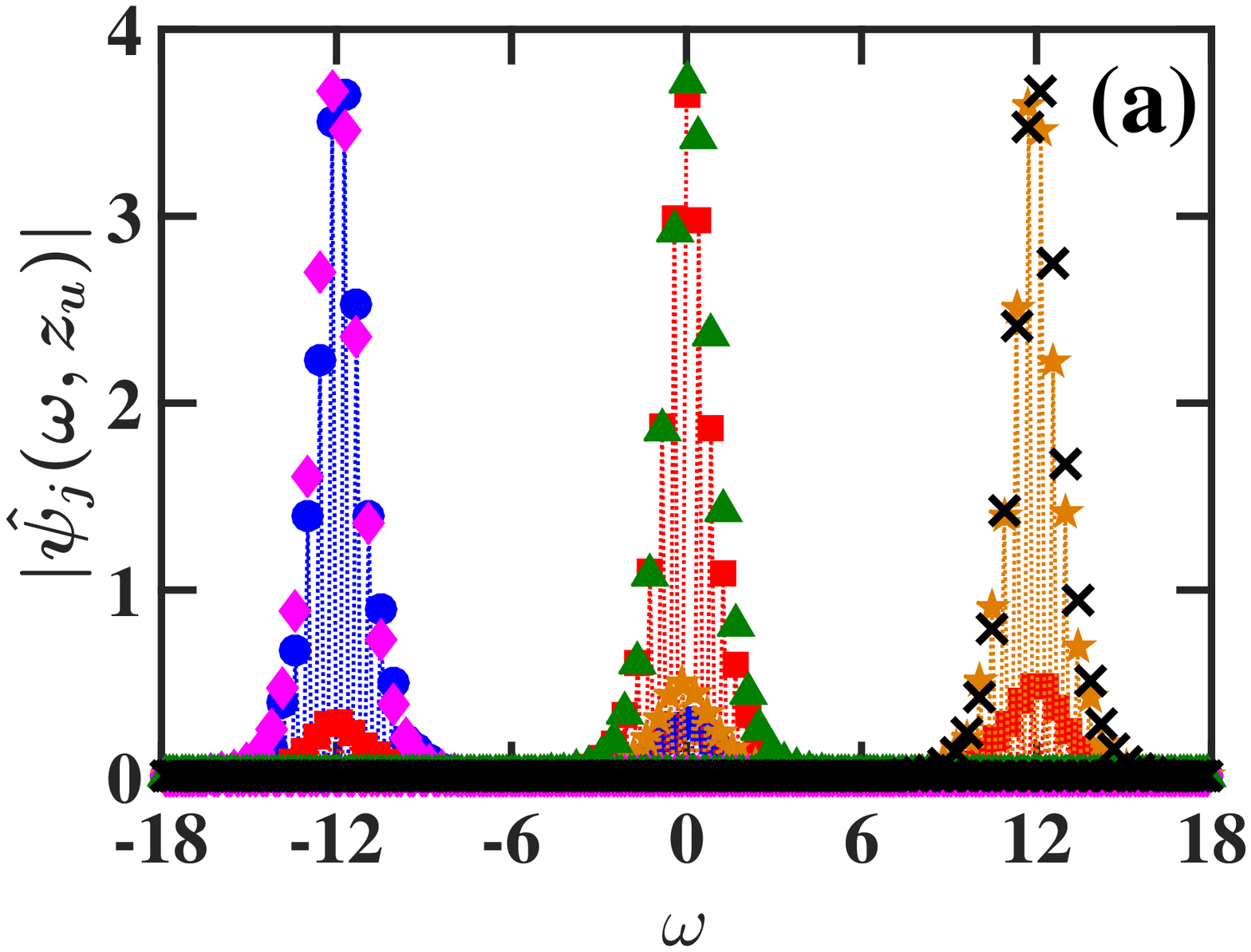} \\
\epsfxsize=11cm  \epsffile{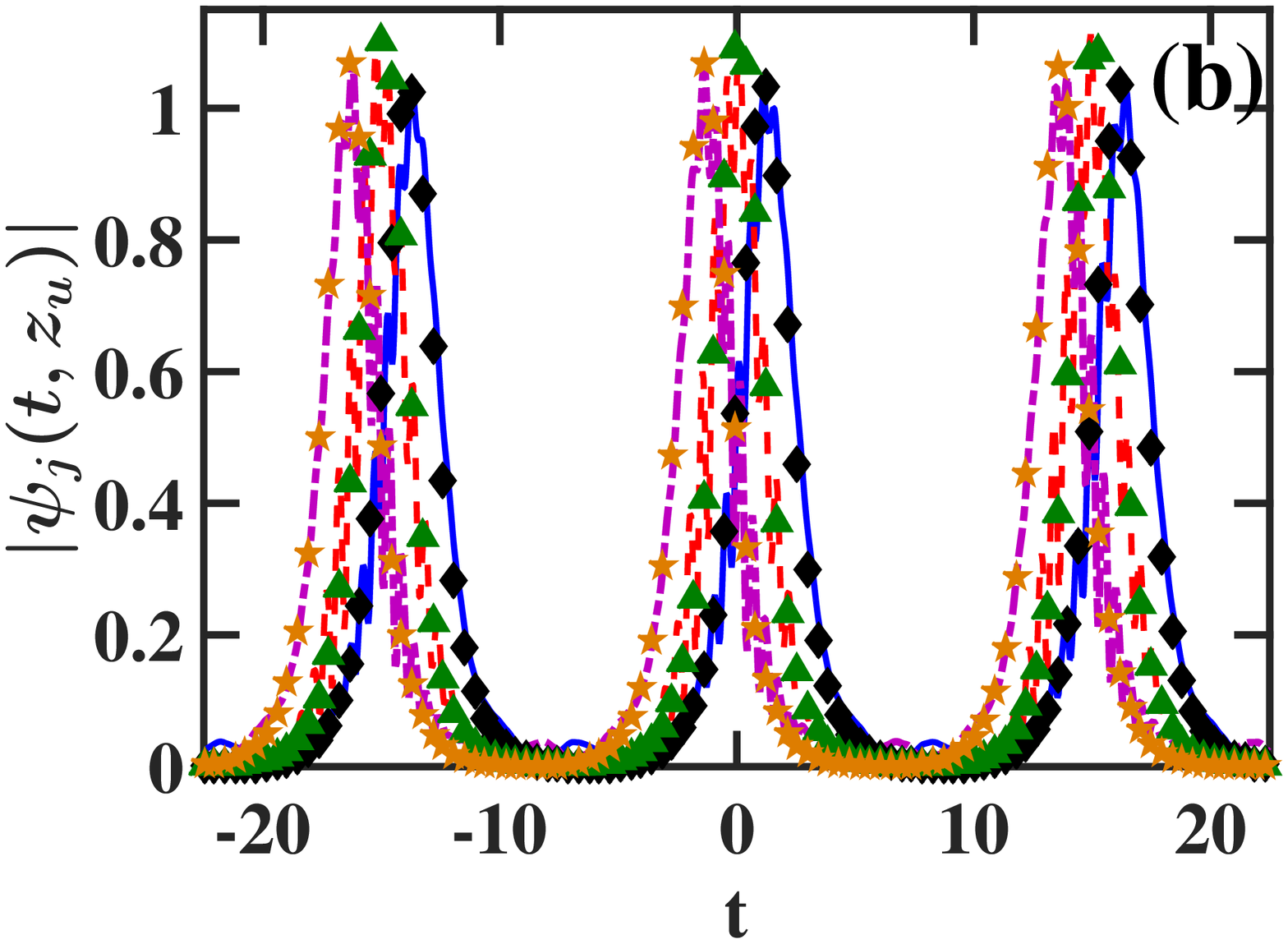} 
\end{tabular}
\end{center}
\caption{(a) The Fourier transforms of the soliton patterns at the onset 
of instability $|\hat\psi_{j}(\omega,z_{u})|$, where $z_{u}=74$, 
for three-channel transmission in a single lossless waveguide with $\gamma=2$, 
$T=15$, and $\Delta\beta=12$. The blue circles, red squares, and orange stars 
represent $|\hat\psi_{j}(\omega,z_{u})|$ with $j=1,2,$ and $3$,
obtained by numerical solution of Eq. (\ref{Kerr1}), 
while the magenta diamonds, green triangles and black crosses correspond 
to the theoretical prediction.        
(b) The soliton patterns at the onset of instability 
$|\psi_{j}(t,z_{u})|$ for three-channel transmission with  
the same parameters used in (a). 
The solid blue, dashed red, and dash-dot purple lines 
correspond to $|\psi_{j}(t,z_{u})|$ with $j=1,2,$ and $3$, 
obtained by the simulations, while the black diamonds, green triangles 
and orange stars correspond to the theoretical prediction.}
\label{fig7}
\end{figure}

It is important to investigate whether the results obtained in subsection \ref{2_channel} 
for transmission stabilization in a two-channel system remain valid as more frequency 
channels are added. For this purpose, we turn to discuss the results 
of numerical simulations for three-channel transmission, starting with 
transmission in a single lossless waveguide. 
As seen in Fig. \ref{fig7}, transmission destabilization is caused by 
resonant formation of radiative sidebands in a manner similar to  
the two-channel case. Moreover, the largest radiative sidebands of the $j$th sequence 
appear at frequencies $\beta_{k}(0)$ of the neighboring soliton sequences.  
That is, the largest sideband of the $j=1$ sequence is formed at 
frequency $\beta_{2}(0)$, the $j=2$ sidebands are formed at frequencies 
$\beta_{1}(0)$ and $\beta_{3}(0)$, and the $j=3$ sideband is formed at 
frequency $\beta_{2}(0)$. Similar to the two-channel case, 
the formation of the radiative sidebands leads to pulse distortion, 
which first appears as fast oscillations in the soliton patterns. 
The growth of the radiative sidebands with increasing propagation distance eventually 
leads to the destruction of the soliton patterns. Furthermore, the distances $z_{u}$,  
at which instability first appears in three-channel transmission, are significantly shorter 
compared with the distances $z_{u}$ observed for two-channel transmission. 
For example, for parameter values $\gamma=2$, $T=15$, and $\Delta\beta=12$, 
used in Figs. \ref{fig1} and \ref{fig7}, $z_{u}=74$ for $N=3$ 
compared with $z_{u}=470$ for $N=2$.                       
 
\begin{figure}[ptb]
\begin{tabular}{cc}
\epsfxsize=11cm  \epsffile{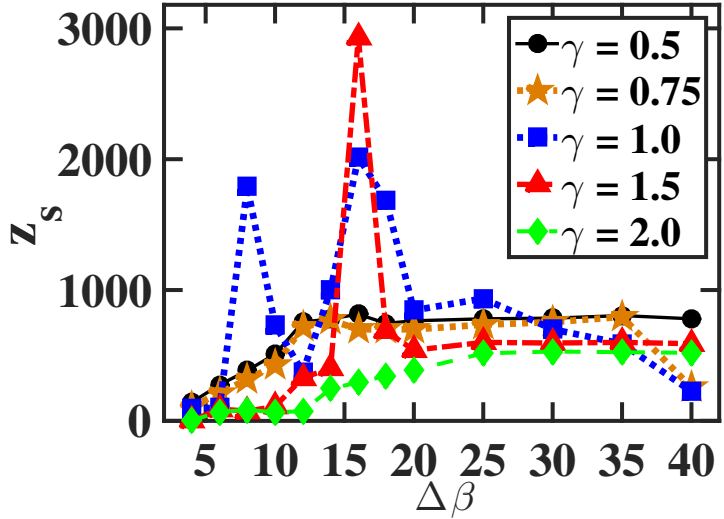} 
\end{tabular}
\caption{Stable propagation distance $z_{s}$ vs frequency spacing 
$\Delta\beta$ for three-channel transmission in a single lossless waveguide 
with $T=15$. The black circles, orange stars, blue squares, red triangles, 
and green diamonds represent the results obtained 
by the simulations for $\gamma=0.5, 0.75, 1.0, 1.5$, and 2.0, respectively.}        
\label{fig8}
\end{figure}

Next, we discuss the dependence of transmission stability for
three-channel transmission in a single lossless waveguide on 
the frequency spacing $\Delta\beta$ and the Kerr nonlinearity coefficient $\gamma$. 
Figure \ref{fig8} shows the stable propagation distances $z_{s}$ 
as functions of the frequency spacing $\Delta\beta$ for 
$T=15$ and $\gamma=0.5, 0.75, 1.0, 1.5$, $2.0$. 
It is observed that the largest $z_{s}$ values are obtained 
with $\gamma=0.5$ for $4 \le \Delta\beta \le 6$, 
$11 \le \Delta\beta < 13$, and $25 < \Delta\beta \le 40$;
with $\gamma=1.0$ for $6 < \Delta\beta < 11$,
$13 \le \Delta\beta < 15$, and $17 \le \Delta\beta \le 25$;  
and with $\gamma=1.5$ for $15 \le \Delta\beta < 17$. 
Thus, there is no single value of $\gamma$, which is 
optimal over the entire frequency spacing interval $4 \le \Delta\beta \le 40$. 
This behavior is sharply different from the one observed for two-channel 
transmission, where the value $\gamma=1.0$ is found to be optimal  
over the entire interval $4 \le \Delta\beta \le 40$. 
Furthermore, the $z_{s}$ values obtained for $N=3$ are significantly smaller 
than the ones obtained for $N=2$. For example, for $20 \le \Delta\beta \le 40$ 
the $z_{s}$ values for three-channel transmission are smaller than 1000 for 
all $\gamma$ values, while the  corresponding $z_{s}$ values 
for two-channel transmission are equal to 5000 for $\gamma=1.0, 1.5$, and $2.0$.

\begin{figure}[ptb]
\begin{center}
\begin{tabular}{cc}
\epsfxsize=11.0cm  \epsffile{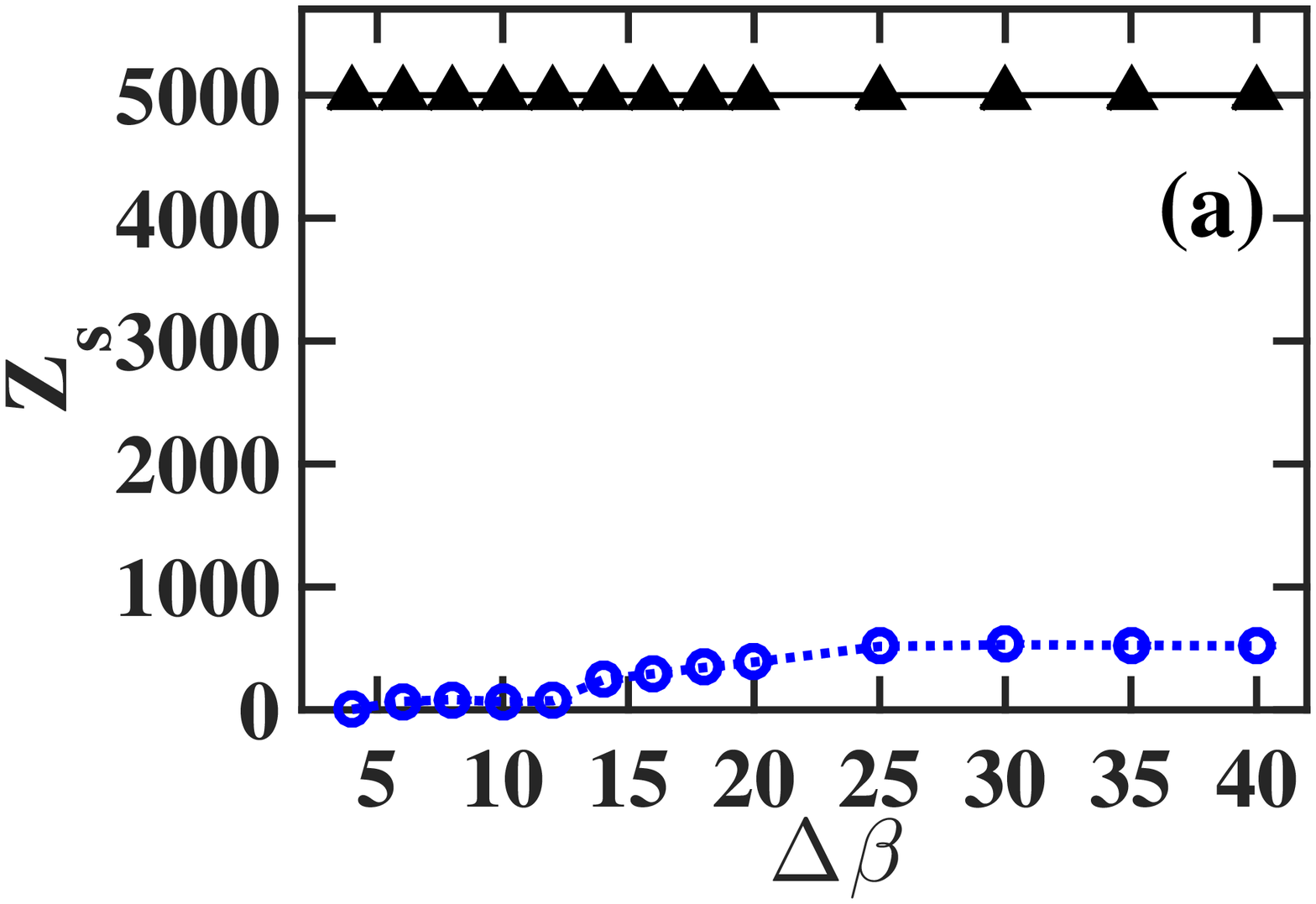}\\
\epsfxsize=11.0cm  \epsffile{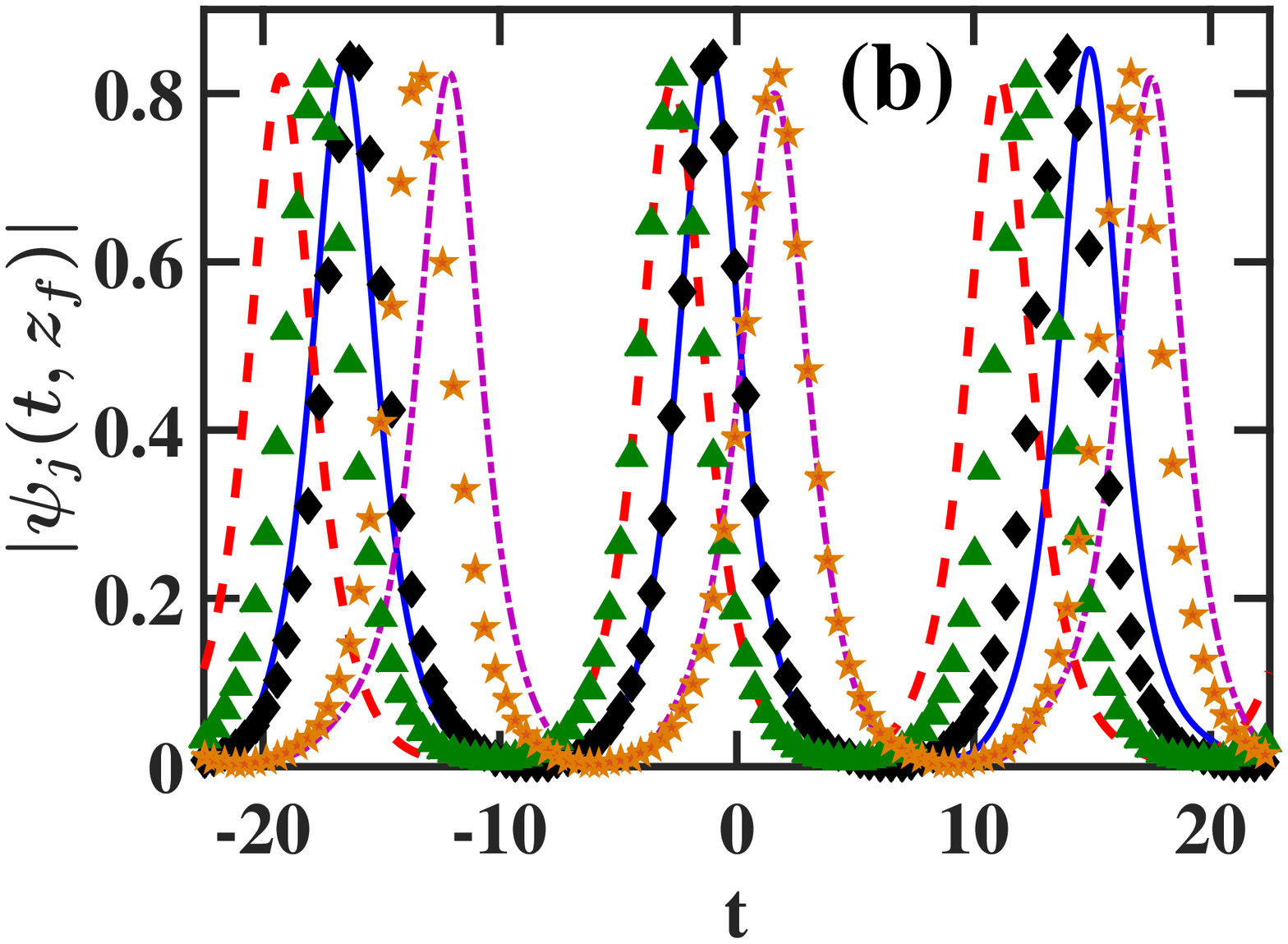}
\end{tabular}
\end{center}
\caption{(a) Stable propagation distance $z_{s}$ vs frequency 
spacing $\Delta\beta$ for three-channel waveguide coupler transmission with 
frequency dependent linear loss and $\gamma=2$, $T=15$, 
$g_{L}=0.5$, $g_{eq}=0$, $\rho=10$, and $W=\Delta\beta/2$.   
The solid black line is the result obtained by 
numerical solution of Eqs. (\ref{Kerr1}) and (\ref{Kerr2}).   
The dashed blue line is the result obtained by the 
simulations for three-channel transmission in a single lossless 
waveguide with $\gamma=2$ and $T=15$.       
(b) The final pulse patterns $|\psi_{j}(t,z_{f})|$, 
where $z_{f}=5000$, in three-channel waveguide coupler transmission 
with $\Delta\beta=12$. The symbols are the same as in Fig. \ref{fig7}(b).}        
\label{fig9}
\end{figure}

We now turn to analyze three-channel transmission in a waveguide coupler 
with frequency dependent linear loss. Our goal is to check whether the 
introduction of frequency dependent linear loss in a waveguide coupler 
leads to enhancement of transmission stability in three-channel systems. 
For this purpose, we numerically solve Eqs. (\ref{Kerr1}) and (\ref{Kerr2}) 
with $N=3$, $g_{eq}=0$, and $g_{L}=0.5$ for $4 \le \Delta\beta \le 40$. 
To enable comparison with the results obtained for two-channel transmission, 
we present here the results of simulations with the same physical parameter 
values as the ones used in Figs. \ref{fig3} and \ref{fig4}. That is, we use 
$\gamma=2$,  $T=15$, $\rho=10$, and $W=\Delta\beta/2$.     
Figure \ref{fig9}(a) shows the stable propagation distance $z_{s}$ 
vs frequency spacing $\Delta\beta$ as  obtained in the simulations 
together with the values obtained for transmission in a single lossless waveguide. 
We observe that $z_{s}=z_{f}=5000$ for all $\Delta\beta$ values 
in the interval $4 \le \Delta\beta \le 40$. Additionally, as seen in Fig. \ref{fig9}(b), 
pulse-pattern distortion is relatively small at the final propagation distance. 
Based on these observations we conclude that three-channel transmission through 
the waveguide coupler is stable throughout the propagation for any 
$\Delta\beta$ value in the interval $4 \le \Delta\beta \le 40$. 
Surprisingly, the $z_{s}$ values obtained for three-channel waveguide coupler 
transmission for $4 \le \Delta\beta \le 7$  are larger than the corresponding 
values obtained for two-channel waveguide coupler transmission by 
factors ranging between 6.45 for $\Delta\beta=4$ and 1.25 for $\Delta\beta=7$.      
Furthermore, the $z_{s}$ values obtained 
for three-channel waveguide coupler transmission are larger than the values 
obtained for three-channel single waveguide transmission  by factors 
ranging between 1250 for $\Delta\beta=4$ and 9.43 for $\Delta\beta=30$. 
Note that these enhancement factors are significantly larger than the enhancement 
factors for two-channel transmission, which are smaller than 172.3 
for all $\Delta\beta$ values in the interval $4 \le \Delta\beta \le 40$.  

\begin{figure}[ptb]
\begin{center}
\begin{tabular}{cc}
\epsfxsize=6.5cm  \epsffile{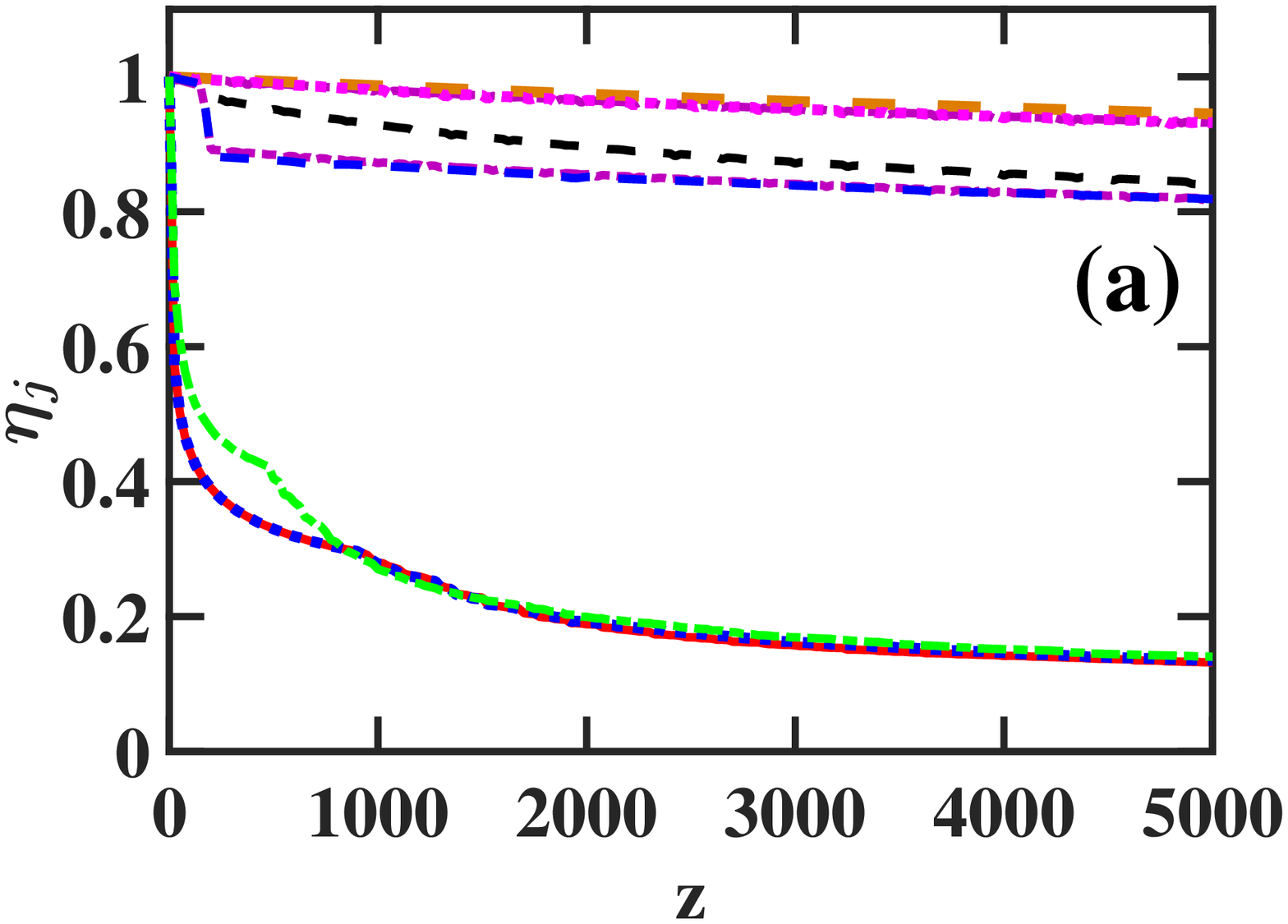} \\
\epsfxsize=6.5cm  \epsffile{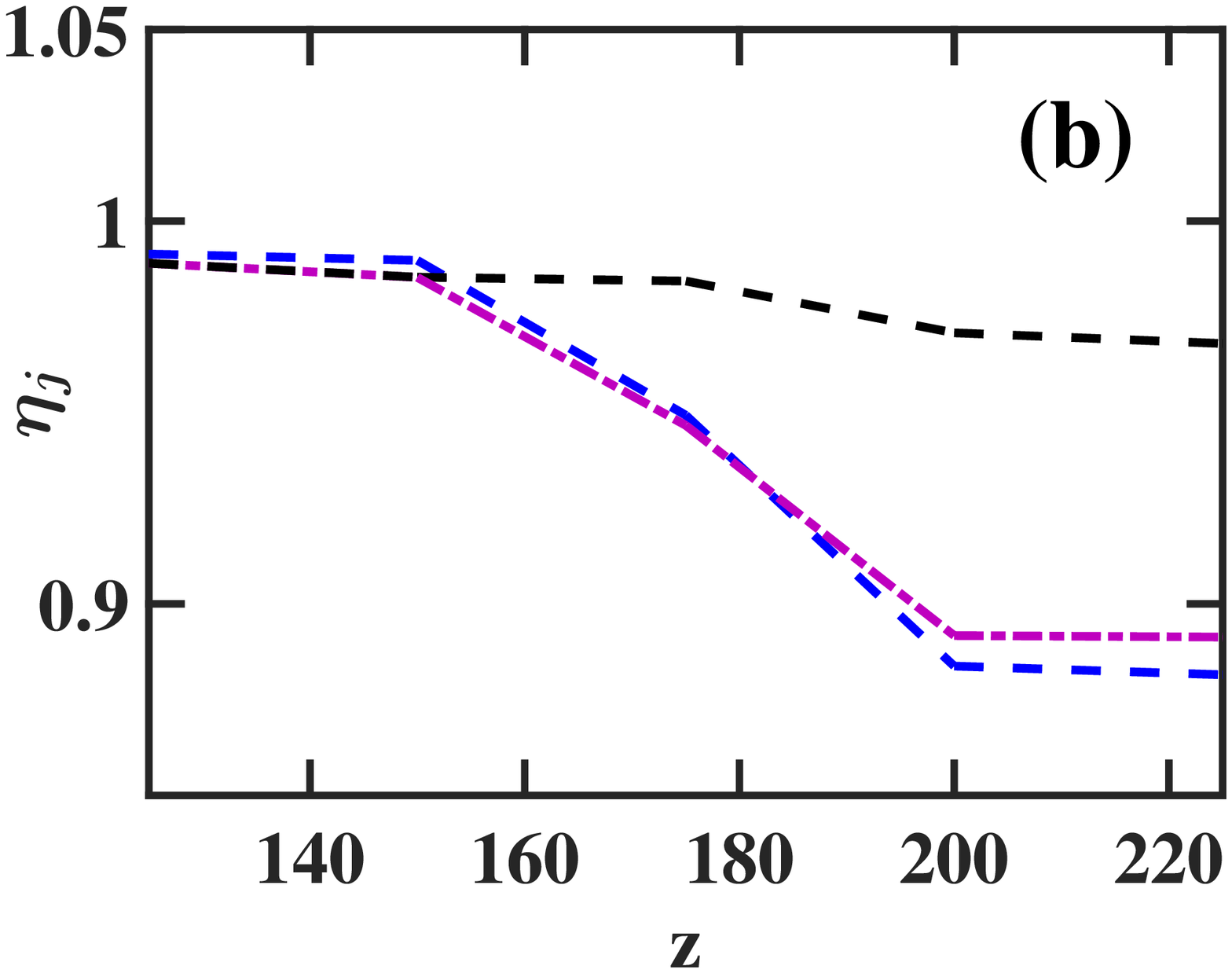} \\
\epsfxsize=6.5cm  \epsffile{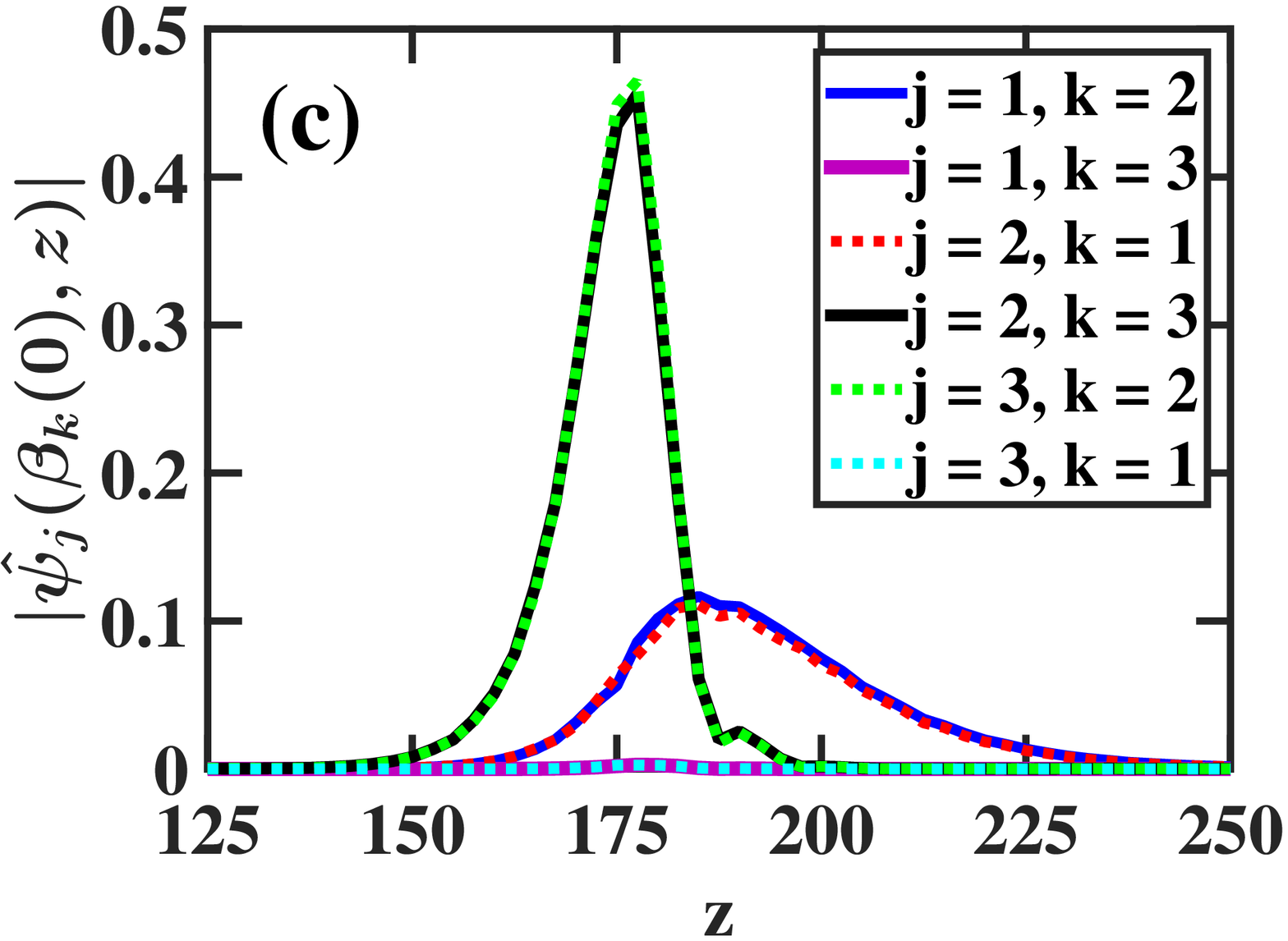}
\end{tabular}
\end{center}
\caption{ (a) The $z$ dependence of soliton amplitudes $\eta_{j}$ 
for three-channel waveguide coupler transmission with frequency dependent 
linear loss and $\gamma=2$, $T=15$, $g_{L}=0.5$, $g_{eq}=0$, 
$\rho=10$, and $W=\Delta\beta/2$.  
The solid red, dashed black, and solid purple curves correspond 
to $\eta_{1}(z)$ obtained by numerical simulations with 
Eqs. (\ref{Kerr1}) and (\ref{Kerr2}) for $\Delta\beta=4$, 
$\Delta\beta=12$, and $\Delta\beta=14$. 
The dashed-dotted-dotted green, short dashed blue, and short 
dashed-dotted orange curves represent $\eta_{2}(z)$ obtained 
by the simulations for $\Delta\beta=4$, $\Delta\beta=12$, 
and $\Delta\beta=14$. The dotted blue, dashed-dotted purple, 
and dotted green curves represent $\eta_{3}(z)$ obtained 
by the simulations for $\Delta\beta=4$, $\Delta\beta=12$, 
and $\Delta\beta=14$. 
(b) Magnified versions of the $\eta_{j}(z)$ 
curves for $\Delta\beta=12$ in the interval $125 \le z \le 250$. 
(c) The $z$ dependence of radiative sideband amplitudes 
$|\hat\psi_{1}(\beta_{2}(0),z)|$ (solid blue line), 
$|\hat\psi_{1}(\beta_{3}(0),z)|$ (solid purple line), 
$|\hat\psi_{2}(\beta_{1}(0),z)|$ (dashed red line), 
$|\hat\psi_{2}(\beta_{3}(0),z)|$ (solid black line), 
$|\hat\psi_{3}(\beta_{1}(0),z)|$ (dashed light blue line), 
and $|\hat\psi_{3}(\beta_{2}(0),z)|$ (dashed green line)
obtained by the simulations for $\Delta\beta=12$.   }        
\label{fig10}
\end{figure}

Further insight into the mechanisms leading to transmission stabilization 
in waveguide couplers with frequency dependent linear loss is gained by 
analyzing the $z$ dependence of soliton amplitudes. 
Similar to the two-channel case, we find three qualitatively different dependences 
of soliton amplitudes on $z$ in the frequency spacing intervals $4 \le \Delta\beta <8$, 
$8 \le \Delta\beta <14$, and $\Delta\beta \ge 14$. 
Figure \ref{fig10}(a) shows the $\eta_{j}(z)$ 
curves obtained by the simulations for three representative $\Delta\beta$ values, 
$\Delta\beta=4$, $\Delta\beta=12$, and $\Delta\beta=14$. 
We observe that for $\Delta\beta=4$ and $\Delta\beta=14$,  
the soliton amplitudes gradually decrease to their final values.  
For $\Delta\beta=12$, the amplitudes of the solitons in the first 
frequency channel also decrease gradually throughout the propagation. 
However, the amplitudes of the solitons in the second and third frequency channels 
exhibit a more complicated dependence on $z$, which is very similar to 
the one observed for two-channel waveguide coupler transmission with $\Delta\beta=12$. 
More specifically, soliton amplitudes in the second and third 
channels gradually decrease for $0 \le z < 150$, but then undergo 
a steep decrease in the interval $150 \le z \le 200$, followed 
by another gradual decrease for $200 < z  \le 5000$ 
[see Figures \ref{fig10}(a) and \ref{fig10}(b)].

\begin{figure}[ptb]
\begin{center}
\begin{tabular}{cc}
\epsfxsize=11cm  \epsffile{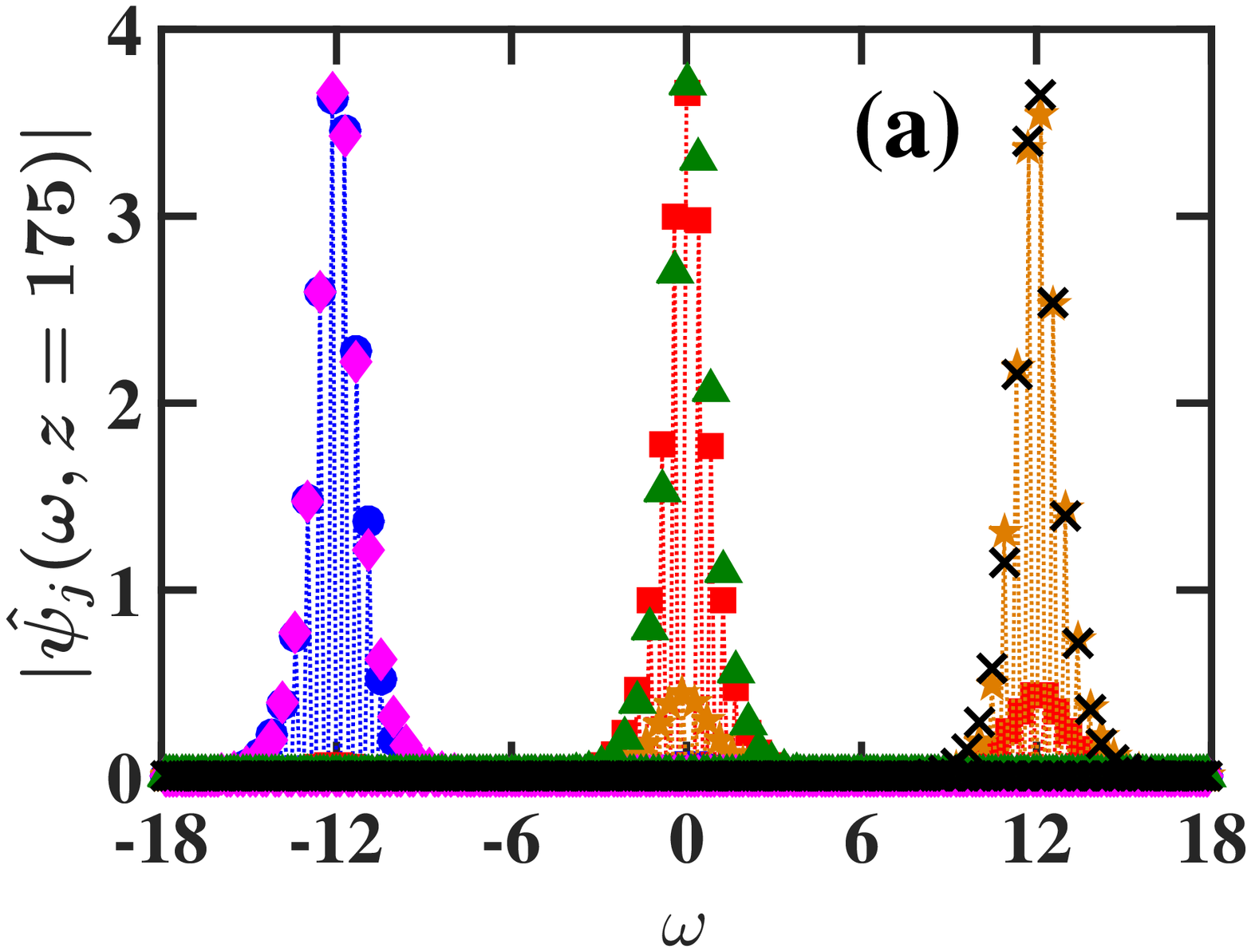} \\
\epsfxsize=11cm  \epsffile{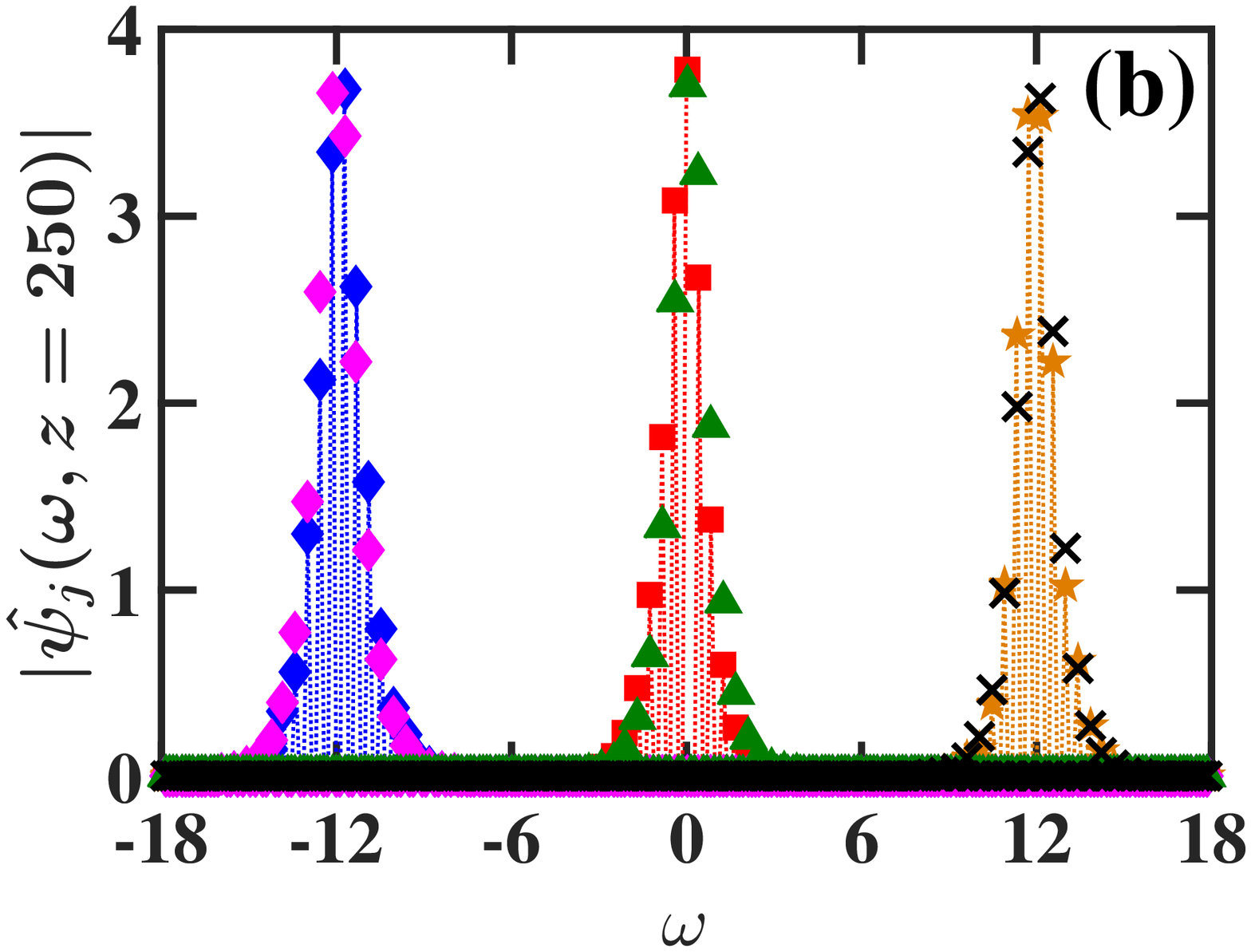} 
\end{tabular}
\end{center}
\caption{The Fourier transforms of the soliton patterns $|\hat\psi_{j}(\omega,z)|$ 
in three-channel waveguide coupler transmission with frequency dependent linear loss 
for $\Delta\beta=12$ and the same values of $\gamma$, $T$, $g_{L}$, 
$g_{eq}$, $\rho$, and $W$ as in Fig. \ref{fig10}. 
(a) $|\hat\psi_{j}(\omega,z)|$ at $z=175$. 
(b) $|\hat\psi_{j}(\omega,z)|$ at $z=250$. 
The symbols are the same as in Fig. \ref{fig7}(a)}        
\label{fig11}
\end{figure}

The behavior of $\eta_{j}(z)$ in the interval $150 \le z \le 200$ 
can be explained by analyzing the $z$ dependence of 
radiative sideband amplitudes $|\hat\psi_{j}(\beta_{k}(0),z)|$, 
where $1 \le j \le 3$, $1 \le k \le 3$, and $j \ne k$. 
Figure  \ref{fig10}(c) shows the $z$ dependence of the radiative sideband amplitudes 
in the interval $125 \le z \le 250$, while Fig.  \ref{fig11} shows 
the Fourier transforms of the soliton patterns at $z=175$ and $z=250$. 
As can be seen from these figures, the sideband amplitudes 
$|\hat\psi_{2}(\beta_{3}(0),z)|$ and $|\hat\psi_{3}(\beta_{2}(0),z)|$ 
attain a sharp maximum at $z=177.5$ with maximal values of 
$0.456$ and $0.465$, respectively. 
The increase of these sideband amplitudes is followed by a drop to 
values smaller than $10^{-3}$ at $z=205$. The formation and subsequent 
decay of the main radiative sidebands for the $j=2$ and $j=3$ channels
in the interval $150 \le z \le 200$ explains the sharp drop in 
$\eta_{2}(z)$ and $\eta_{3}(z)$ observed in this interval. 
Indeed, the formation of the sidebands leads to energy transfer from a 
localized soliton form to a nonlocalized radiative form, which results in 
the steep drop of $\eta_{2}(z)$ and $\eta_{3}(z)$. Additionally, 
the strong linear loss $g_{L}$ at frequencies $\beta_{3}(0)$ for 
$j=2$ and $\beta_{2}(0)$ for $j=3$ leads to the relatively fast decay 
of the sidebands following their formation.    
We note that the evolution of $\eta_{2}(z)$ and $\eta_{3}(z)$ in the three-channel 
waveguide coupler is in fact quite similar to the evolution of 
$\eta_{1}(z)$ and $\eta_{2}(z)$ in the two-channel waveguide coupler.

Figure  \ref{fig10}(c) also shows that the sideband amplitudes 
$|\hat\psi_{1}(\beta_{2}(0),z)|$ and $|\hat\psi_{2}(\beta_{1}(0),z)|$ 
attain a maximum at $z=185$ with maximal values of $0.117$ and 
$0.111$, respectively. The increase of these sideband amplitudes is 
followed by a decrease to below $10^{-3}$ values at $z=257.5$. 
Thus, the formation and subsequent decay of the $j=1$ sideband at 
frequency $\beta_{2}(0)$ in an interval centered about $z=185$ 
explains the observed drop in the value of $\eta_{1}(z)$ in this interval. 
Additionally, the sideband amplitudes $|\hat\psi_{1}(\beta_{3}(0),z)|$ 
and $|\hat\psi_{3}(\beta_{1}(0),z)|$ attain a maximum at $z=177.5$, 
but the corresponding maximal values are smaller than $10^{-3}$, 
and as a result, do not significantly affect the amplitude dynamics. 
The relatively small values of $|\hat\psi_{1}(\beta_{3}(0),z)|$ 
and $|\hat\psi_{3}(\beta_{1}(0),z)|$ 
compared with the other four sideband amplitudes indicate 
that the magnitude of radiative sidebands decreases 
as the absolute value of the frequency difference 
$|\beta_{j}(0)-\beta_{k}(0)|$ increases.

  
\begin{figure}[ptb]
\begin{center}
\begin{tabular}{cc}
\epsfxsize=11.0cm  \epsffile{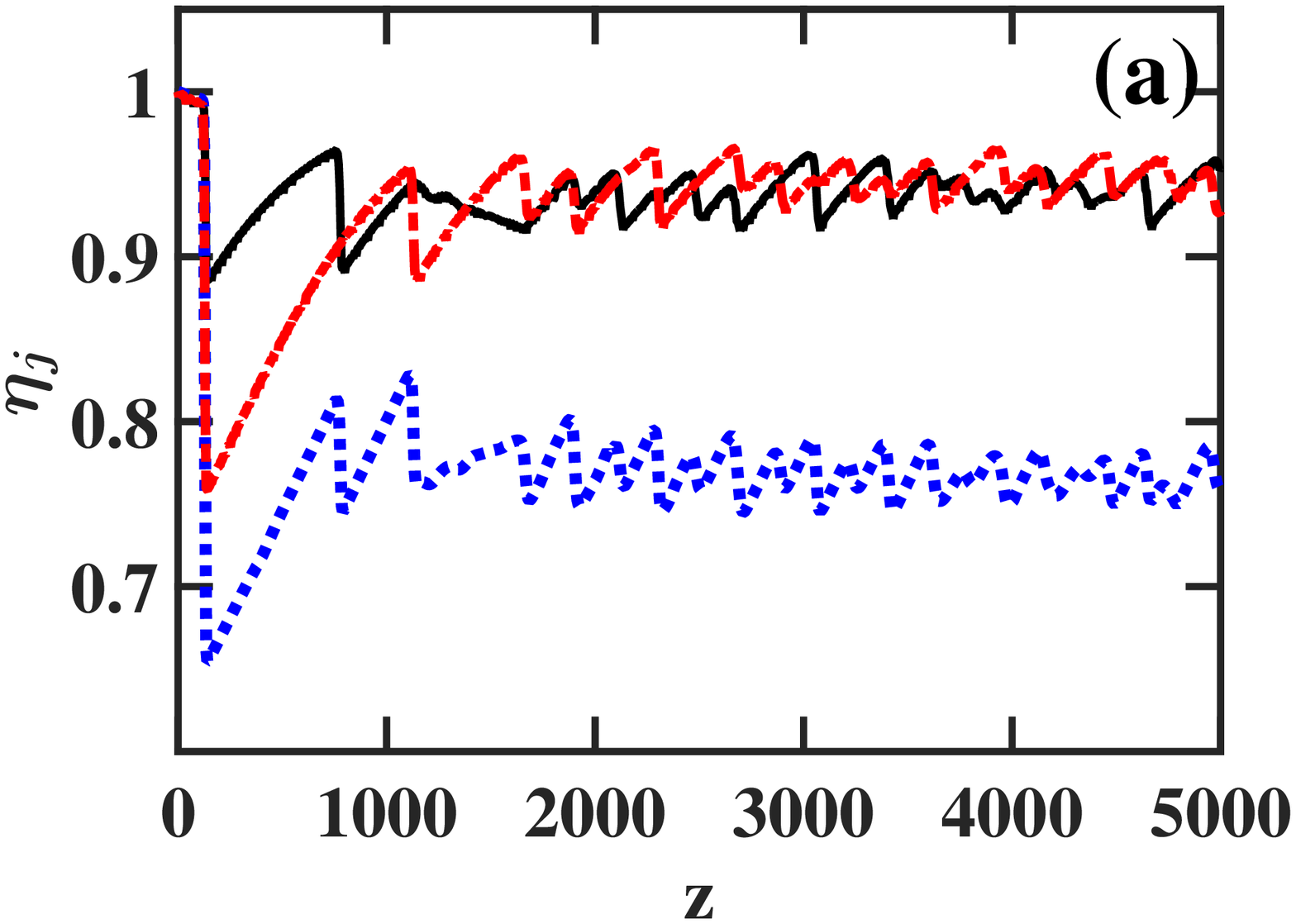} \\
\epsfxsize=11.0cm  \epsffile{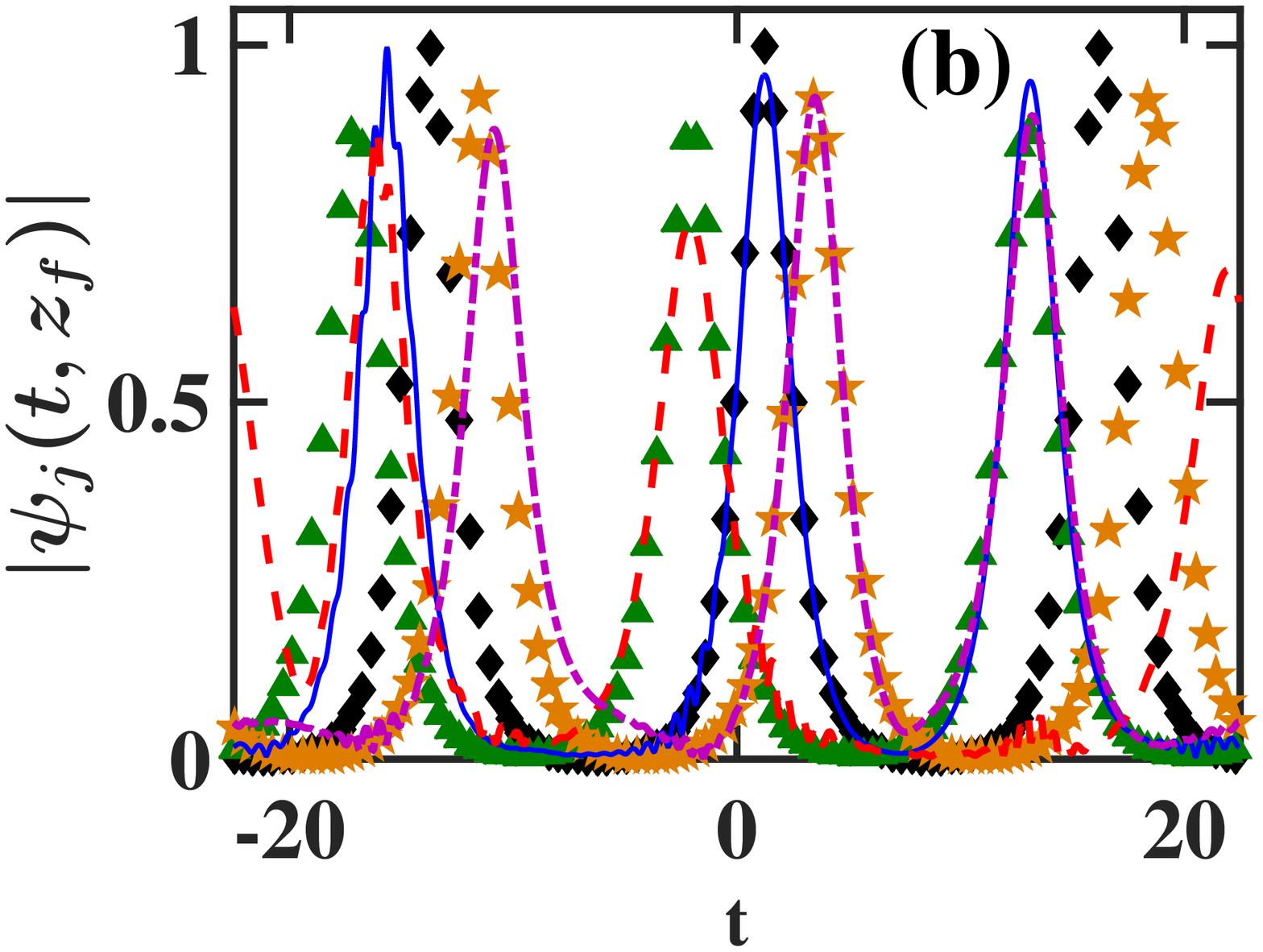}
\end{tabular}
\end{center}
\caption{(a) The $z$ dependence of soliton amplitudes $\eta_{j}$ 
for three-channel waveguide coupler transmission with frequency dependent linear 
gain-loss and $\gamma=2$, $T=15$, $\Delta\beta=10$,   
$g_{L}=0.5$, $g_{eq}=3.9 \times 10^{-4}$, $\rho=10$, and $W=5$. 
The solid black, dashed blue, and dashed-dotted red lines 
correspond to $\eta_{j}(z)$ with $j=1,2, 3$, 
as obtained by numerical solution of Eqs. (\ref{Kerr1}) and (\ref{Kerr2}). 
(b) The final pulse patterns $|\psi_{j}(t,z_{f})|$, where $z_{f}=5000$.    
The symbols are the same as in Fig. \ref{fig7}(b).}
\label{fig12}
\end{figure}

We conclude the discussion of three-channel transmission by considering 
propagation in waveguide couplers with frequency dependent linear gain and loss. 
As explained in section \ref{CNLS}, in these waveguide couplers, 
the weak linear gain $g_{eq}$ in 
the frequency interval $(\beta_{j}(0)-W/2, \beta_{j}(0)+W/2]$
is expected to enable soliton propagation without amplitude decay. 
To check if such stable propagation can indeed be realized, 
we numerically solve Eqs.  (\ref{Kerr1}) and (\ref{Kerr2}) 
with $N=3$ and with a value of $g_{eq}$, which is determined by Eq.  (\ref{Kerr2C}). 
To enable comparison with the results presented in Fig. \ref{fig6} for two-channel 
transmission, we discuss the results of numerical simulations with the same set of 
physical parameter values. That is, we use $\gamma=2$, $T=15$, $\Delta\beta=10$, 
$g_{eq}=3.9 \times 10^{-4}$, $g_{L}=0.5$, $\rho=10$, and $W=5$. 
Figure \ref{fig12}(a) shows the $z$ dependence of soliton amplitudes obtained 
by the simulations. It is seen that the amplitudes undergo a sharp drop, which is followed by 
oscillations about values of $\eta_{s1}=0.940$, $\eta_{s2}=0.768$, 
and $\eta_{s3}=0.947$, for $j=1$, $j=2$ and $j=3$, respectively.  
In addition, as seen in Fig. \ref{fig12}(b), 
the soliton shape is retained at $z_{f}=5000$, although the pulses in 
each sequence experience significant position shifts relative to one another. 
We note that $\eta_{s2}$ is significantly smaller than $\eta_{s1}$ and $\eta_{s3}$. 
In addition, the overall oscillatory dynamics of soliton amplitudes is similar to the one 
observed in two-channel transmission, although the pattern of oscillations is more 
complex for $N=3$ compared with $N=2$.

\begin{figure}[ptb]
\begin{center}
\begin{tabular}{cc}
\epsfxsize=7.2cm  \epsffile{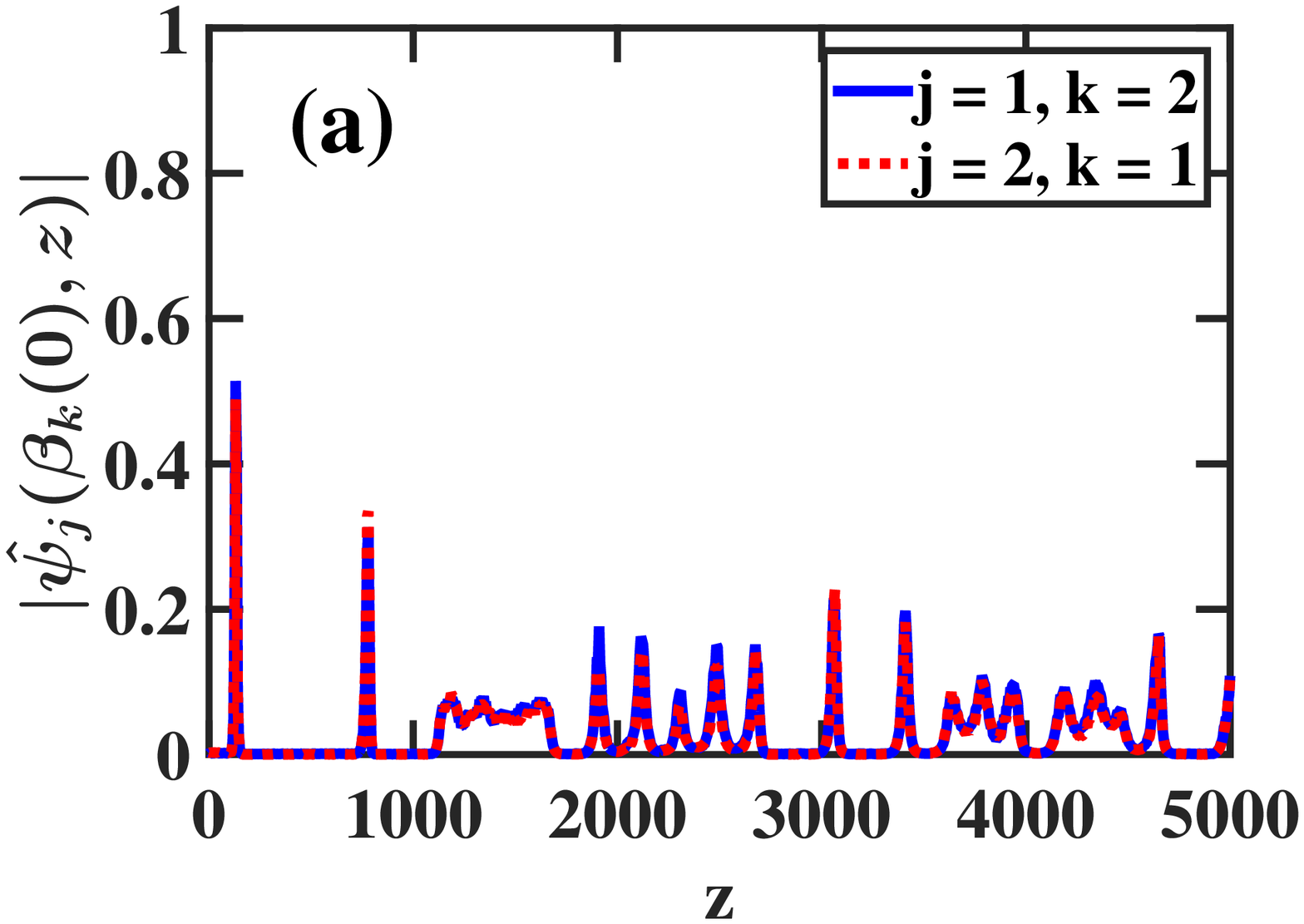}\\
\epsfxsize=7.2cm  \epsffile{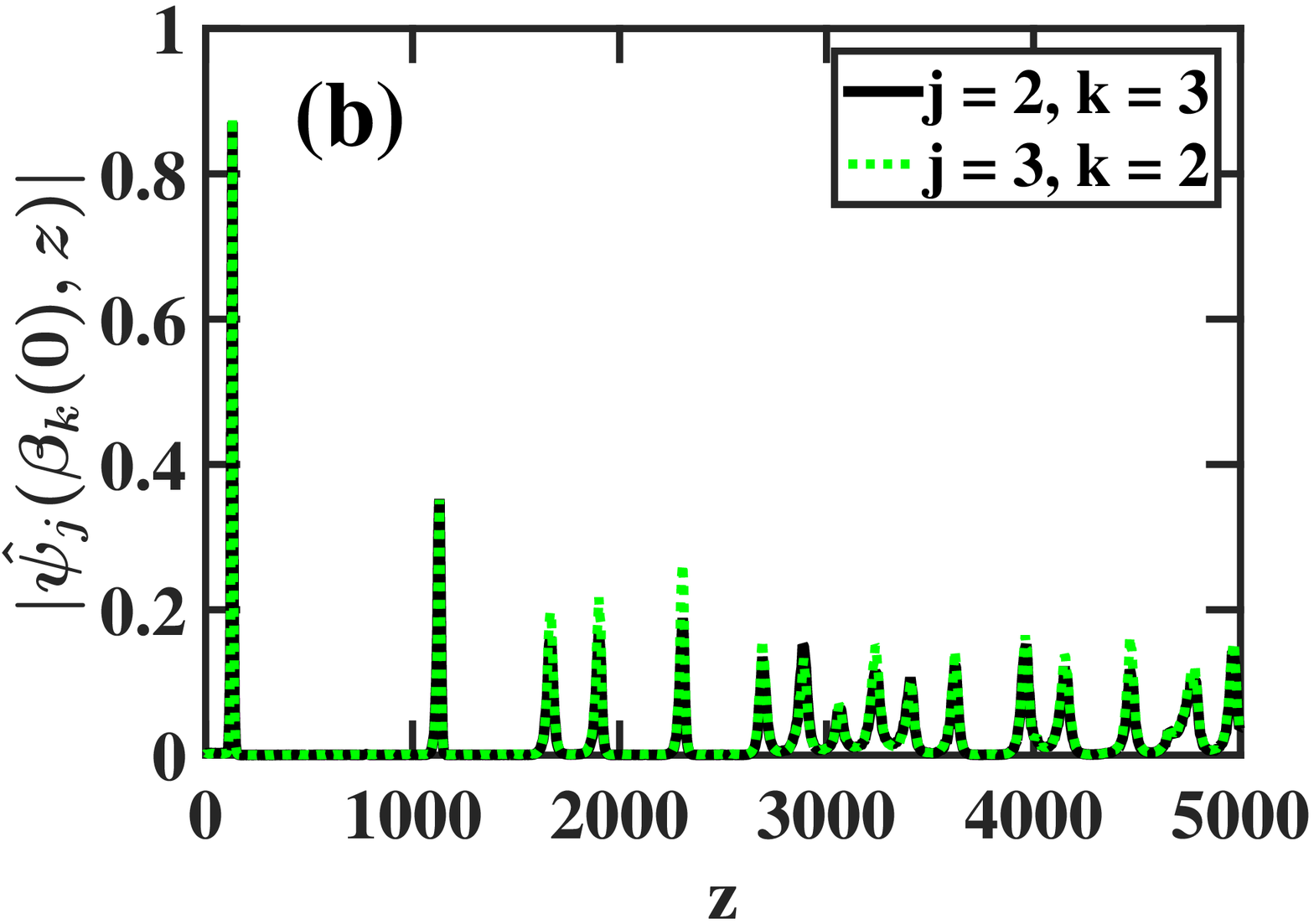}\\
\epsfxsize=7.2cm  \epsffile{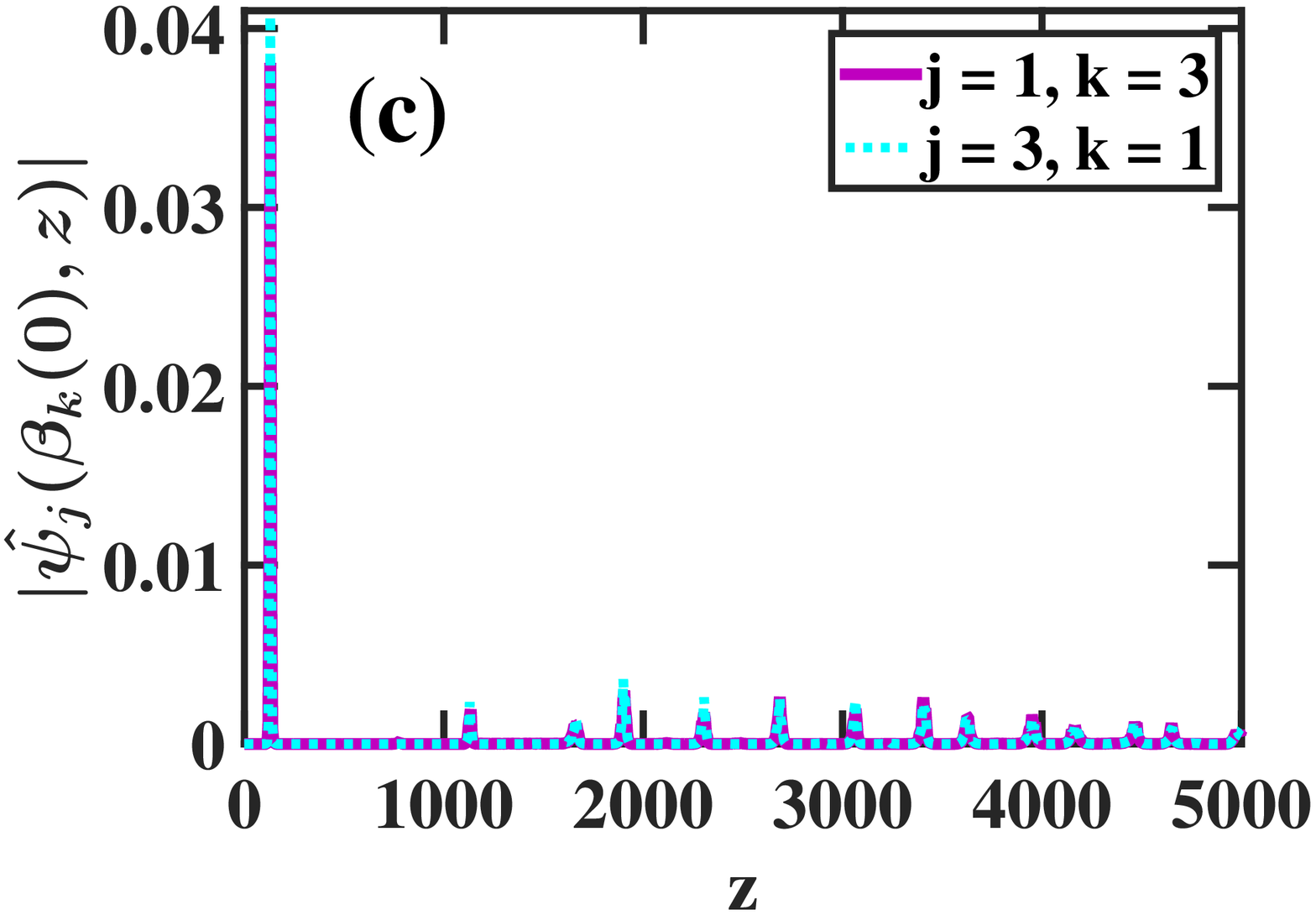}
\end{tabular}
\end{center}
\caption{The $z$ dependence of radiative sideband amplitudes 
for three-channel waveguide coupler transmission with 
the same physical parameter values as in Fig. \ref{fig12}. 
(a) $|\hat\psi_{1}(\beta_{2}(0),z)|$ and $|\hat\psi_{2}(\beta_{1}(0),z)|$ vs z. 
(b) $|\hat\psi_{2}(\beta_{3}(0),z)|$ and $|\hat\psi_{3}(\beta_{2}(0),z)|$ vs z. 
(c) $|\hat\psi_{1}(\beta_{3}(0),z)|$ and $|\hat\psi_{3}(\beta_{1}(0),z)|$ vs z. 
All curves represent results obtained by simulations with Eqs. (\ref{Kerr1}) and (\ref{Kerr2}). 
The symbols are the same as in Fig. \ref{fig10}(c).}        
\label{fig13}
\end{figure}

Similar to the situation in two-channel transmission, the oscillations of 
soliton amplitudes can be related to radiative sideband dynamics. 
We study this dynamics by analyzing the $z$-dependence of  
sideband amplitudes $|\hat\psi_{j}(\beta_{k}(0),z)|$,  
where $1 \le j \le 3$, $1 \le k \le 3$, and $j \ne k$. 
Figure \ref{fig13}(a) shows the 
$z$ dependence of sideband amplitudes $|\hat\psi_{1}(\beta_{2}(0),z)|$ 
and $|\hat\psi_{2}(\beta_{1}(0),z)|$, Fig. \ref{fig13}(b) shows the $z$ 
dependence of $|\hat\psi_{2}(\beta_{3}(0),z)|$ and $|\hat\psi_{3}(\beta_{2}(0),z)|$, 
while Fig. \ref{fig13}(c) shows the $z$ dependence of $|\hat\psi_{1}(\beta_{3}(0),z)|$ 
and $|\hat\psi_{3}(\beta_{1}(0),z)|$. All curves in Fig. \ref{fig13} 
are obtained by numerical solution of Eqs. (\ref{Kerr1}) and (\ref{Kerr2}). 
We note that the values of 
$|\hat\psi_{1}(\beta_{3}(0),z)|$ and $|\hat\psi_{3}(\beta_{1}(0),z)|$ 
are smaller than 0.041 throughout the propagation, 
and therefore these sidebands do not significantly affect amplitude dynamics.  
We therefore focus attention on dynamics of the four strongest sidebands 
$|\hat\psi_{1}(\beta_{2}(0),z)|$, $|\hat\psi_{2}(\beta_{1}(0),z)|$, 
$|\hat\psi_{2}(\beta_{3}(0),z)|$, and $|\hat\psi_{3}(\beta_{2}(0),z)|$. 
As seen in Figure \ref{fig13}, the radiative sideband amplitudes experience alternating 
``periods'' of growth and decay, similar to the situation in two-channel transmission.  
Furthermore, the distances at which sideband amplitudes attain their maxima 
for the four strongest sidebands are located inside the relatively short intervals, 
where soliton  amplitudes are decreasing. 
Therefore, the dynamics of the radiative sidebands 
can indeed be related to the oscillatory dynamics of soliton amplitudes. 
More specifically, as the sidebands grow, energy is transferred from 
a localized soliton form to a nonlocaized form, resulting in a 
decrease in soliton amplitudes. 
The strong linear loss $g_{L}$ outside the central frequency intervals 
leads to a relatively fast decay of the sidebands, and as a result, 
the sidebands maxima are very narrow with respect to $z$. 
Additionally, the weak linear gain $g_{eq}$ at the central frequency intervals 
leads to the slow growth of soliton amplitude at the subsequent waveguide spans
and to the overall oscillatory dynamics.

Figure \ref{fig13} also provides an explanation for the smaller value 
of $\eta_{s2}$ compared to $\eta_{s1}$ and $\eta_{s3}$. 
Indeed, during the first (and largest) drop in soliton amplitudes, 
the solitons in the $j=2$ sequence lose energy due to 
formation of radiative sidebands at {\it both} $\beta_{1}(0)$ and $\beta_{3}(0)$. 
In contrast, the $j=1$ and $j=3$ solitons lose energy almost entirely due 
to formation of radiative sidebands at $\beta_{2}(0)$, since the sidebands 
at $\beta_{3}(0)$ for $j=1$ and at $\beta_{1}(0)$ for $j=3$ are very small. 
In addition, the complicated pattern of radiative sideband growth and decay 
shown in Fig. \ref{fig13} is responsible for the more complex pattern 
of amplitude oscillations observed in three-channel transmission compared 
with two-channel transmission [compare Fig. \ref{fig12}(a) with Fig. \ref{fig6}(a) 
and Fig. \ref{fig13} with Fig. \ref{fig6}(c)].

\section{Conclusions}
\label{conclusions}
In summary, we made several major theoretical steps towards realizing stable long-distance 
multichannel soliton transmission in Kerr nonlinear waveguide loops. 
We found that transmission destabilization in a single lossless waveguide 
is caused by resonant formation of radiative sidebands
due to intersequence cross-phase modulation. 
We then showed that in two-channel systems,  
significant enhancement of the stable propagation distance, 
which holds over a wide range of interchannel frequency spacing values,  
is obtained by optimization with respect to the Kerr nonlinearity coefficient $\gamma$. 
In contrast, we found that in three-channel transmission in a single lossless waveguide, 
no single value of the Kerr nonlinearity coefficient is optimal 
for the entire interval of interchannel frequency spacings that we examined. 
Moreover, we developed a general method for transmission stabilization, 
based on frequency dependent linear gain-loss in 
Kerr nonlinear waveguide couplers, and implemented the method 
in two-channel and three-channel transmission.  
We showed that the introduction of frequency 
dependent loss leads to significant enhancement of 
transmission stability even for non-optimal $\gamma$ values 
via decay of radiative sidebands, which can be described as a dynamic phase transition. 
For waveguide couplers with frequency dependent linear gain-loss, 
we observed stable oscillations of soliton amplitudes 
due to decay and regeneration of radiative sidebands.     
Transmission stabilization was achieved without  
dispersion-management or filtering.  

\section*{Acknowledgments}
Q.M.N. is supported by the Vietnam National Foundation for Science and Technology 
Development (NAFOSTED). 
D.C. is grateful to the Mathematics Department of NJCU 
for providing technological support for the computations.


\begin{thebibliography}{}
\bibitem{dimensions} The dimensionless distance $z$ in Eq. (\ref{Kerr1})  is 
$z=X/(2L_{D})$, where $X$ is the dimensional distance, 
$L_{D}=\tau_{0}^{2}/|\tilde\beta_{2}|$ is the dimensional dispersion length,    
$\tau_{0}$ is the soliton width, and $\tilde\beta_{2}$ is the second-order
dispersion coefficient. The dimensionless time is
$t=\tau/\tau_{0}$, where $\tau$ is the time. 
$\psi_{j}=E_{j}/\sqrt{P_{0}}$, where $E_{j}$ is proportional to the 
electric field of the $j$th sequence and $P_{0}$ is the peak power.   
The coefficients $\gamma$ and $g_{eq}$ are related to the
dimensional Kerr nonlinearity and linear gain-loss coefficients 
$\tilde\gamma$ and $\tilde g_{eq}$ by    
$\gamma=2P_{0}\tau_{0}^{2}\tilde\gamma/|\tilde\beta_{2}|$ 
and $g_{eq}=2\tau_{0}^{2}\tilde g_{eq} /|\tilde\beta_{2}|$, respectively. 
The solitons spectral width is $\nu_{0}=1/(\pi^{2}\tau_{0})$ and the 
intersequence frequency difference is $\Delta\nu=(\pi\Delta\beta\nu_{0})/2$.
\end{thebibliography}
\end{document}